\pdfoutput=1
\documentclass[twocolumn,reprint,superscriptaddress,footinbib,aps,prb]{revtex4-2}

\usepackage[mathscr]{euscript}

\usepackage{graphicx}

\newcommand{\subfigref}[2]{\ref{#1}\hyperref[#1]{\textbf{(#2)}}}

\usepackage{amsmath,physics,dsfont}

\usepackage[x11names,table]{xcolor}

\usepackage[]{hyperref}
\hypersetup{
    colorlinks,
    linkcolor={IndianRed3},
    citecolor={Cyan4},
    urlcolor={Cyan4}
}

\def\d{\hat{d}}
\def\dd{\hat{d}^\dagger}
\def\b{\hat{b}}
\def\bb{\hat{b}^\dagger}
\def\gw{\lambda}
\def\c{\hat{c}_{k, \nu}}
\def\cc{\hat{c}^\dagger_{k,\nu}}
\def\t{{t}_{k,\nu}}
\def\tt{{t}^*_{k,\nu}}
\def\eps{\epsilon_{k,\nu}}
\def\G{\Upsilon_{\nu}}
\def\f{f_{\nu}}
\def\D{\hat{D}}
\def\DD{\hat{D}^\dagger}
\def\tHs{\tilde{H}_s}
\def\tVr{\tilde{V}}
\def\tVrI{\tilde{V}_{I}}
\def\Tr{\text{Tr}}
\def\R{\text{R}}
\newcommand{\E}[2]{\tilde{E}_{{#1},{#2}}}
\newcommand{\SZeroSOneD}[4]{\delta_{{#2},0}\delta_{{#4},1} \DD_{{#1} {#3} }}
\newcommand{\SOneSZeroD}[4]{\delta_{{#2},1}\delta_{{#4},0} \D_{{#1} {#3} }}
\newcommand{\e}[2]{\omega_{{#1},{#2}}}
\newcommand{\Rnn}[6]{\text{A}^{{#1}{#2}}_{{#3}{#4}\,{#5}{#6}}\Big|_\nu}

\def\IL{I_L}
\def\IR{I_R}
\def\I{I_\nu}
\def\S{\tilde{S}}
\def\B{\tilde{B}}
\def\trho{\tilde{\rho}}

\def\rhoR{\rho_\text{R}}
\newcommand{\Int}[1]{{#1}_I}
\newcommand{\Pbra}[1]{\langle {#1}|} 
\newcommand{\Pket}[1]{|{#1}\rangle } 
\newcommand{\Fbra}[2]{\langle {#1},{#2}|} 
\newcommand{\Fket}[2]{|{#1},{#2}\rangle } 
\def\NESS{non-equilibrium steady state}
\def\qd{QD}
\def\resonator{QHO}

\begin{document}

\title{Quantum master equation for nanoelectromechanical systems beyond the wide-band limit}

\author{Sofia Sevitz}
\affiliation{University of Potsdam, Institute of Physics and Astronomy, Karl-Liebknecht-Str. 24-25, 14476 Potsdam, Germany\looseness=-1}

\author{Federico Cerisola}
\affiliation{Physics and Astronomy, University of Exeter, Exeter EX4 4QL, United Kingdom}

\author{Janet Anders}
\affiliation{University of Potsdam, Institute of Physics and Astronomy, Karl-Liebknecht-Str. 24-25, 14476 Potsdam, Germany\looseness=-1}
\affiliation{Physics and Astronomy, University of Exeter, Exeter EX4 4QL, United Kingdom}

\date{\today} 


\begin{abstract}

Coupling the vibrations of an oscillator to electronic transport is a key building block for nanoelectromechanical systems. They describe many nanoscale electrical components such as molecular junctions. Inspired by recent experimental developments, we derive a quantum master equation that describes nanoelectromechanical systems in a generally overlooked situation: when the electronic transport is slower than the natural frequency of the oscillator. Here, a semi-classical model is no longer valid and we develop the missing fully quantum approach. Moreover, we go beyond the wide-band limit and study the consequence of maintaining energy dependent tunneling rates, which are required to describe effects found in real devices. To benchmark our results, we compare with numerically exact results obtained with the hierarchical equations of motion method, and find overall good agreements in the experimentally accessible steady state regime. Furthermore, we derive from the microscopic model a ready to use particle current expression that replicates features already observed experimentally.

\end{abstract}

\maketitle


\section{Introduction}\label{Sec:Introduction}

\begin{figure}[t]
    \centering
    \includegraphics[width=\linewidth]{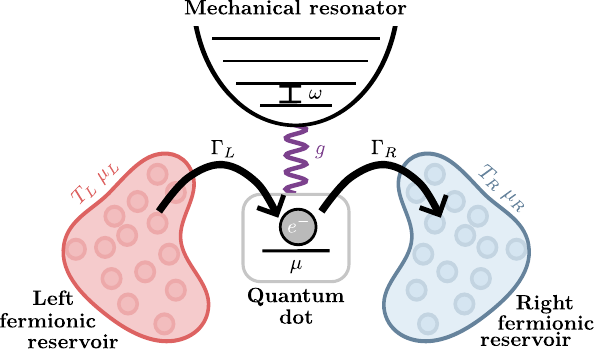}
    \caption{
    \textbf{Illustration of a nanoelectromechanical system.}
    A quantum dot with chemical potential $\mu$ is coupled (with arbitrary strength $g$) to a quantum harmonic oscillator with natural frequency $\omega$. Moreover, the quantum dot is coupled to two fermionic reservoirs to the left ($L$) and right ($R$) with temperatures $T_L$ and $T_R$, and chemical potentials $\mu_L$ and $\mu_R$, respectively. A temperature and/or chemical imbalance between the reservoirs enables hopping of electrons from one reservoir to the other (via the quantum dot) with electronic tunneling rates $\Gamma_{L/R}$.
    }
    \label{fig:Figure1}
\end{figure}

Nanoelectromechanical systems (NEMS) are characterized by the intrinsic coupling of electronic transport to mechanical motion~\cite{Craighead2000, Blencowe2005}. Typical set-ups include a quantum dot (\qd) in contact with fermionic reservoirs that hold a temperature and/or chemical imbalance, allowing for electronic tunneling to induce a net current flow. In turn, the transported electron is coupled to a mechanical degree of freedom, typically modeled as a simple quantum harmonic oscillator (\resonator), see Fig.~\ref{fig:Figure1}. This simple system covers a wide-range of platforms ranging from carbon nanotubes with an embedded quantum dots~\cite{Vigneau2022,Luo2017,Meerwaldt2012,Steele2009,Samanta2023,Laird2012,Moser2014,Schmid2015,Tabanera-bravo2024} to molecular junctions~\cite{Leijnse2010, Mitra2004, Erpenbeck2018}. Due to their favorable properties such as high quality factors~\cite{Huttel2009}, a new wave of interest has been directed towards the use of these nanoscale devices in a variety of fields, spanning from metrology~\cite{Chaste2012} to thermodynamics~\cite{De2016}. 

Given the typical experimentally relevant time scales~\cite{Luo2017, Sevitz2025, Wen2020}, the main regime of interest for these platforms is the \NESS. In the search to model this state, it is customary to consider two main theoretical approximations: (i) electronic transport is significantly faster than the oscillations of the resonator, known as the \textit{fast tunneling regime} or the \textit{quasiadiabatic} limit; and (ii) energy-independent electronic tunneling rates, namely the \textit{wide-band} limit. The first approximation allows for a semi-classical model, which results in a classical Fokker-Planck~\cite{Wachtler2019} or Langevin equation~\cite{Culhane2022} for the \resonator~motion and classical rate equations to describe the electronic transport~\cite{Vigneau2022}. The second approximation significantly simplifies the analytical treatment of the transport properties, allowing for closed expressions for the currents~\cite{Schaller2014Book}. These common approximations have been widely employed to model devices that operate within their regime of validity~\cite{Culhane2022, De2016, Clerk2005, Pistolesi2007, Micchi2015, Vigneau2022, Luo2017, Barnard2019, Wang2021, Wen2020}. However, some nanomechanical platforms, such as suspended carbon nanotubes, can achieve very high mechanical frequencies~\cite{Laird2012}, so that electron tunneling can be tuned to a regime slower than the characteristic oscillator time. Then, the semi-classical approach (i) is no longer valid, and a different approach to a tractable model is needed. Moreover, there is experimental evidence of the need of going beyond the energy-independent electronic tunneling rates assumption (ii) to properly reproduce certain observed effects~\cite{Meerwaldt2012, Tabanera-bravo2024}.

In this work, we derive a master equation to model NEMS~in the \textit{slow tunneling regime}, i.e. when the electronic transport is slower than the frequency of the \resonator. Moreover, our master equation captures the effect of energy dependent tunnel barriers. In these ways, we provide an analytical model to describe NEMS in a regime not covered by commonly employed approaches. We restrict our analysis to the single electron tunneling regime, which is well captured by the second order perturbation master equation~\cite{Timm2008}. While in most of the article we focus on the \NESS~of the system, we note that our model can also describe the transient dynamics. To benchmark our master equation, we compare with the numerically exact solutions as calculated by the quantum Hierarchical Equations of Motion (HEOM) method~\cite{SchinabeckThesis2019} and find overall good agreements. Finally, we derive a ready to use expression for the electronic particle current. Comparing with previous experimental results, we obtain that the computed currents showcase expected behavior.


\section{The model beyond wide-band}

In this section we describe the model considered and the consequence of maintaining the energy dependence on the tunneling rates, going beyond approximation (ii). Throughout the article we set $\hbar=k_B=1$. The Hamiltonian
\begin{equation}\label{eq:H_total}
    \hat{H}= \hat{H}_s \otimes \mathds{1}_\R+\mathds{1}_s \otimes \hat{H}_\R + \hat{V}
\end{equation}
describes the entire set-up showcased in Fig.~\ref{fig:Figure1}. In the following we describe each component separately.

The first term details the NEMS~composed of a \qd~with a single energy level $\mu$ coupled to a \resonator~with mechanical frequency $\omega$. The Hamiltonian reads~\cite{Culhane2022,De2016,Mitra2004,Leijnse2010,Piovano2011}
\begin{equation}\label{eq:H_s}
    \hat{H}_s = \mu \dd \d +  \omega \bb \b +  g (\bb+\b)\dd\d.
\end{equation}
On the one hand, $\dd$($\d$) are the creation (annihilation) operator of the dot that follow fermionic statistics. On the other hand, $\bb$($\b$) are the creation (annihilation) operators of the resonator that follow bosonic statistics. The intrinsic coupling considered here is between the electronic population of the dot ($\dd \d$) and the displacement of the resonator ($\bb +\b$), with a coupling strength $g$. In recent carbon nanotube experiments, this coupling was found to be in the ultrastrong regime $\gw=g/\omega \gg 1$, implying a strong back-action that needs to be accounted for~\cite{Vigneau2022,Samanta2023}.

The second term of Eq.~\eqref{eq:H_total} describes the two fermionic reservoirs
\begin{equation}\label{eq:H_R}
    \hat{H}_\R =\sum_{\nu}\sum_k \epsilon_{\nu k} \cc \c, 
\end{equation}
where $\epsilon_{\nu k}$ is the free energy of an electron with wave vector $k$ in the reservoir $\nu=\{L,R\}$. We consider the creation (annihilation) operators $\cc$($\c$) in the reservoirs under the Klein transformation such that they follow fermionic statistics among themselves but commute with the system operators, i.e. $[\d,\c]=0$ and $[\b,\c]=0$ for all $k,\nu$~\cite{Schaller2014Book, Potts2024}\footnote{A similar argument can also be made using the Jordan-Wigner transformation.}. 

These reservoirs can exchange electrons with the \qd~via the dot-reservoir coupling described in the last term of Eq.~\eqref{eq:H_total} as
\begin{align}\label{eq:V_r}
    \hat{V} =\sum_{\nu} \sum_k t_{\nu k} \cc \d  + \mathrm{h.c},
\end{align}
where the coupling amplitudes $t_{\nu k}$ are determined by the tunneling rates $\G(\epsilon)=2\pi\sum_k |t_{\nu k}|^2\delta(\epsilon-\epsilon_{\nu k})$. As mentioned in the introduction, it is customary in the transport community to take the \textit{wide-band} limit, resulting in energy-independent rates, $\G(\epsilon) = \Gamma_\nu$ constant
\cite{Kirsanskas2017, Mahan1996, De2016, Leijnse2010, Mayrhofer2021, Esposito2012, Thijssen2008, Usui2024, Moulhim2021, Potts2021, Erpenbeck2018}. While this commonly employed approximation is sufficient for many situations, it is also known to have fundamental shortcomings~\cite{Wang2010,Cheng2024}. In particular, for NEMS the wide-band approximation fails to capture key effects such as self-oscillations~\cite{Blanter2004, Clerk2005, Wachtler2019, Culhane2022}. Here, we go beyond this approximation and maintain the energy-dependence $\G(\epsilon)$ which more accurately describes experimental platforms~\cite{Meerwaldt2012, Tabanera-bravo2024}. We consider a centered peaked spectral density around $\gamma_\nu$ with broadening $\delta_\nu$, which we model by a single Lorentzian peak~\cite{Erpenbeck2018_2,Schaller2013}
\begin{equation}\label{eq:spectral_Lorentz}
    \G(\epsilon)=\frac{\Gamma_\nu \delta_\nu^2}{(\epsilon-\gamma_\nu)^2 +\delta_\nu^2},
\end{equation} 
with $\Gamma_\nu$ the dot-reservoir tunneling rate. The \textit{wide-band} limit is recovered by taking $\delta_\nu \rightarrow \infty$. Our work can be easily extended to a general spectral density constructed by a superposition of Lorentz shapes with shifted centers, we refer the reader to the Appendix~\ref{Appendix:Lamb_Shift} for more details.
\begin{figure}[t]
    \centering
    \includegraphics[width=1\linewidth]{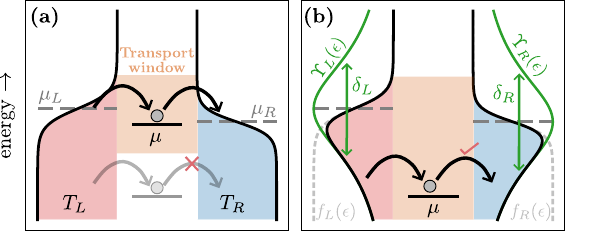}
    \caption{\textbf{Illustrative example to showcase the physical consequence of going beyond the wide-band limit.}
    \textbf{(a)} Energy spectrum of the dot and the two reservoirs under the wide-band limit ($\delta_L,\delta_R \rightarrow \infty$ in Eq.~\eqref{eq:spectral_Lorentz}). The shaded area denotes the fermionic occupation corresponding to the left reservoir (red) and right reservoir (blue). The shaded orange area denotes the transport window of the dot, that is the energy range where electronic transport between the reservoirs is enabled. Outside this window, the electronic occupation prohibits the \textit{hopping} of electrons.
    \textbf{(b)} Same as in panel \textbf{(a)} but with tunneling rates beyond the wide-band limit with a finite width of $\delta_L=\delta_R = 50$ GHz in the spectral densities given by Eq.~\eqref{eq:spectral_Lorentz} (green). For reference, in dashed gray, the fermi functions are indicated. Note that the transport window is energy broadened compared to panel \textbf{(a)} thus allowing for electronic transport in energies previously prohibited. 
    Parameters: $\Gamma_L/2\pi=\Gamma_R/2\pi=0.2$ GHz, $T_L= 6.546$ (corresponding to $50$ mK), $T_L= 5.237$ (corresponding to $40$ mK), $\mu_L=-\mu_R= 5$ GHz.
    }
    \label{fig:Figure2}
\end{figure}
Maintaining the energy dependence of the reservoir coupling has 
a direct impact on particle tunneling probabilities between the dot and the
leads. Indeed, these are given by the $in\,(0\rightarrow 1)/out\,(1\rightarrow 0)$
electronic transition rates defined as
\begin{align}
    \label{def:QD_R01}
    R^{0\rightarrow 1}_\nu(\epsilon) &= \G(\epsilon) \f(\epsilon) \\
    \label{def:QD_R10}
    R^{1\rightarrow 0}_\nu(\epsilon) &= \G(\epsilon) [1-\f(\epsilon)],
\end{align}
respectively. Here $\f(\epsilon)=1/(\exp(\beta_\nu(\epsilon-\mu_\nu))+1)$ is the Fermi distribution of reservoir $\nu$ with chemical potential $\mu_\nu$ and inverse temperature $\beta_\nu=1/T_\nu$. Contrary to commonly employed expressions for these transition rates~\cite{Vigneau2022,Dorsch2022Thesis}, they now possess an additional energy dependence due to the $\G(\epsilon)$ spectral density. To gain intuition on its effect, we showcase an example in Fig.~\ref{fig:Figure2}. Here, we plot the energy spectrum of the dot and the two reservoirs under the wide-band limit (Fig.~\subfigref{fig:Figure2}{a}) and with a finite spectral width (Fig.~\subfigref{fig:Figure2}{b}). The shaded areas denote the fermionic occupation of the energy levels of the left reservoir (red) and right reservoir (blue). In Fig.~\subfigref{fig:Figure2}{b}, due to the finite width of the Lorentz spectral density, the energy dependence of fermionic occupations of the reservoir is altered; for energies below the chemical potential $\mu_\nu$ the effective electronic function does not plateau to a constant value but instead follows the form of $\G(E)$, in this case the Lorentz function. Now, the electronic states that were completely occupied in the wide-band limit case (Fig.~\subfigref{fig:Figure2}{a}) can now host an additional electron. Naturally, this alteration has an effect on the allowed energies for electrons to \textit{hop} from the left reservoir to the right (or vise versa) known as the dot's transport window (orange shaded area). This will have consequences in the electronic transport, as we will show in Sec.~\ref{Sec:Current}. Finally, we note that maintaining the energy dependence of the tunneling rates is key to be able to describe phenomena such as self-oscillations of the mechanical resonator~\cite{Culhane2022, Sevitz2025Engine}, unlike the more phenomenological approach commonly taken in electron-shuttle models~\cite{Wachtler2019,Wu2025,Utami2006,Pistolesi2005}.

In the following section we derive the master equation 
for the system maintaining the energy-dependence on the tunneling rates under a fully open quantum system approach. 


\section{Non-equilibrium steady state}\label{Sec:NESS}

In this section we detail how to obtain the \NESS~of the system defined as 
\begin{equation}\label{eq:NESS_def}
    \partial_t \rho_s= 0,
\end{equation}
where $\rho_s$ is the reduced NEMS~(\qd+\resonator) density matrix. In principle, the left hand side of Eq.~\eqref{eq:NESS_def} is given by the von Neumann equation 
\begin{equation}\label{eq:rho_s_VN}
    \partial_t \rho_s= -i\, \Tr_\R\Big([\hat{H},\rho]\Big).
\end{equation}
As mentioned in Sec.~\ref{Sec:Introduction}, in this work we only consider the regime where single electron sequential tunneling is valid, i.e. $ \Gamma_\nu \ll T_\nu$. This is captured by the second order perturbation over the dot-reservoir coupling in Eq.~\eqref{eq:rho_s_VN}~\cite{Timm2008,Mayrhofer2021,Schaller2013}. In the following, we apply physically motivated assumptions to the right hand side of Eq.~\eqref{eq:rho_s_VN} in order to obtain a master equation. 


\subsection{Polaron transformation}

Before deriving the master equation, it is convenient to work in the representation in which the system's Hamiltonian $\hat{H}_s$ is diagonal. We do this by taking the polaron transformation defined as~\cite{Nazir2016,Wuger1998,Kolli2011} 
\begin{equation}\label{eq:Polaron_trans}
    \tilde{\bullet}=U \bullet U^\dagger,
\end{equation}
with $U=e^{\gw (\bb-\b) \dd\d}$. Applying Eq.~\eqref{eq:Polaron_trans} to the total Hamiltonian defined in Eq.~\eqref{eq:H_total} yields 
\begin{equation}\label{eq:H_total_polaron}
    \tilde{H}= \tHs \otimes \mathds{1}_\R + \mathds{1}_s \otimes \hat{H}_\R + \tVr.
\end{equation}
Note that the transformation affects the system's Hamiltonian (Eq.~\eqref{eq:H_s}) and the reservoir coupling (Eq.~\eqref{eq:V_r}) while leaving the reservoir Hamiltonian (Eq.~\eqref{eq:H_R}) invariant. For the system, the polaron transformation incorporates the dot-resonator interaction term as a shift in the dot's chemical potential as~\cite{Siddiqui2007,Piovano2011,Schaller2014Book}
\begin{equation}\label{eq:H_s_tilde}
    \tHs=\tilde{\mu} \dd \d  +  \omega \bb \b. 
\end{equation}
with $\tilde{\mu}=\mu-\omega \gw^2$. 

We define the basis $|j,n\rangle=\Pket{j} \otimes |n\rangle$ with $\Pket{j} =\Pket{0}, \Pket{1}, \dots , \Pket{N}$ representing the mechanical phonon Fock states for the \resonator~and $|n\rangle= |0\rangle,|1\rangle$ representing the \qd~occupation states, which diagonalizes Eq.~\eqref{eq:H_s_tilde} as
\begin{equation}\label{eq:Energy_system}
    \tHs \Fket{j}{n} = \tilde{E}_{jn} \Fket{j}{n}
\end{equation}
with eigenenergies $\tilde{E}_{jn}=n \tilde{\mu} + j  \omega$. We reserve the lower case $n$ for the electronic state in the \qd. 

The polaron frame affects the reservoir coupling as~\cite{Siddiqui2007,Piovano2011,Schaller2014Book}
\begin{equation}\label{eq:V_r_tilde}
    \tVr=\sum_{\nu,k} t_{\nu k}  \cc  \d  \hat{D}(-\gw) +h.c.,
\end{equation}
where $\hat{D}(\gw)= e^{\gw (\bb-\b)}$ is the displacement operator for the \resonator. Note that now the reservoir $\nu$ is coupled to both the \qd~and the \resonator~via its momentum.


\subsection{Redfield master equation}\label{Sec:Redfield_main_text}

We now proceed to our main objective of computing the \NESS~as defined in Eq.~\eqref{eq:NESS_def}. In what follows we will motivate the main approximations and state the final expressions we obtain. For a complete derivation we refer the reader to Appendix~\ref{Appendix:ME}. 

We assume that initially ($t_0=-\infty$) the system and the reservoirs are un-correlated~\cite{Timm2008, Mitra2004, Piovano2011} $\trho_{s}(t_0)\otimes \rhoR$, where $\rhoR =  \bigotimes_{\nu} \frac{1}{Z_\nu}\exp\{- \beta_\nu\left(\sum_k (\epsilon_{\nu k}-\mu_\nu) \hat{c}^\dagger_{\nu k}\hat{c}_{\nu k}\right)\}$ is the density matrix of the reservoirs at thermal equilibrium and $\trho_{s}(t_0)$ is the initial state of the NEMS.
Next, we assume that the reservoirs are infinite such that they are not affected by the electronic tunnelings and are always in thermal equilibrium~\cite{Timm2008}. Moreover, we take the Markov approximation, where the system dynamics at time $t$ is determined \textit{only} by the operators at the same time~\cite{Potts2021}. Finally, as here we go beyond the wide-band limit and take the tunneling rates with their energy dependence, as discussed in~\cite{Zedler2009} for the case of a single QD, the additional condition $\delta_\nu\gg \Gamma_\nu$ is needed.

After lengthy calculations, see Appendix~\ref{Appendix:ME}, we obtain the projected Redfield equation given by
\begin{widetext}
\begin{equation}\label{Eq:ME_Redfield_final}
    \Pbra{j}\partial_t\trho_s^n \Pket{m} = -i\omega(j-m)\Pbra{j} \trho_s^n \Pket{m} + \sum_\nu \sum_{kl}
    \Big( -\Rnn{n,}{n}{j}{m}{k}{l} \Pbra{k} \trho_s^n \Pket{l}
    + \Rnn{n+1,}{n}{j}{m}{k}{l} \Pbra{k} \trho_s^{n+1} \Pket{l}
     + \Rnn{n-1,}{n}{j}{m}{k}{l} \Pbra{k} \trho_s^{n-1} \Pket{l} \Big),
\end{equation}
\end{widetext}
where for the dot occupation $n=0,1$ we have made use of the notation $ \Fbra{k}{n} \trho_s \Fket{l}{n} =\Pbra{k} \trho_s^n \Pket{l}$. We only consider the diagonal element of the dot's degree of freedom given that state superposition is not allowed due to the particle super selection rule~\cite{Potts2024, Bartlett2007}. Contrary to previous works~\cite{Culhane2022,Leijnse2010}, as we are in the slow tunneling regime, we allow the \resonator~to develop coherences due to the \qd+\resonator~coupling. 

In Eq.~\eqref{Eq:ME_Redfield_final} we have introduced the \textit{Redfield tensors} corresponding to lead $\nu$ as
\begin{align}\nonumber
    \label{eq:R_0_0}
    \Rnn{0}{0}{j}{m}{k}{l} & =  \frac{1}{2} \sum_i \Big( R^{0\rightarrow 1}_\nu(\e{k}{i})  \DD_{j,i}\D_{i,k} \delta_{ml} \\  & \quad \quad \quad \quad \quad +  R^{0\rightarrow 1}_\nu(\e{l}{i})) \DD_{l,i}\D_{i,m} \delta_{kj} \Big)  \\  
    \label{eq:R_0_1}
    \Rnn{0}{1}{j}{m}{k}{l} &= \frac{1}{2}\D_{j,k}\DD_{l,m}  \Big( R^{0\rightarrow 1}_\nu(\e{k}{j})  + R^{0\rightarrow 1}_\nu(\e{l}{m})\Big)\\ \nonumber
    \label{eq:R_1_1} 
    \Rnn{1}{1}{j}{m}{k}{l} & =  \frac{1}{2} \sum_i 
    \Big( R^{1\rightarrow 0}_\nu(\e{i}{k}) \D_{j,i} \DD_{i,k}\delta_{ml} \\  & \quad \quad \quad \quad \quad + R^{1\rightarrow 0}_\nu(\e{i}{l}) \D_{l,i} \DD_{i,m} \delta_{jk} \Big) \\
    \label{eq:R_1_0}
    \Rnn{1}{0}{j}{m}{k}{l} &= \frac{1}{2} \DD_{j,k}\D_{l,m} \Big(R^{1\rightarrow 0}_\nu(\e{j}{k}) + R^{1\rightarrow 0}_\nu(\e{m}{l}) \Big)
\end{align}
where $\e{k}{l}=\tilde{\mu}-\omega(k-l)$, $\D_{kl} = \Pbra{k}e^{\gw (\bb-\b)} \Pket{l}$ and $R^{0\rightarrow 1}_\nu/R^{1\rightarrow 0}_\nu$ are the electronic transition rates previously defined in Eq.~\eqref{def:QD_R01}-\eqref{def:QD_R10}. 

Eq.~\eqref{Eq:ME_Redfield_final} constitutes the main result of this article. In Appendix~\ref{Appendix:QD_recovery} we show that if one takes no coupling among dot and resonator, i.e. $\gw=0$, then we recover the standard rate equations for a \qd. Although related results have been reported in Ref.~\cite{Mitra2004} and~\cite{Piovano2011}, to our knowledge, this is the first time this derivation has been made in the slow tunneling regime, making use of a open quantum system approach, and without use of the wide-band limit. This last consideration is a crucial point as recent experiments evidence that the energy dependence on the tunneling rates cannot be neglected~\cite{Meerwaldt2012,Tabanera-bravo2024}. As a final remark, we note that Eq.~\eqref{Eq:ME_Redfield_final} neglects the Lamb shift. We refer the reader to Appendix~\ref{Appendix:Lamb_Shift} for further details. 

In the following section we benchmark the sates obtained with the master equation derived here with the exact solutions obtained using HEOM~\cite{Erpenbeck2018_2,Huang2023}. Although we focus on the \NESS, the transient dynamics can, of course, also be calculated, and we will come back to this in a later section.


\subsection{Benchmark of \NESS}\label{Sec:Benchmark}

To benchmark the \NESS~obtained from Eq.~\eqref{Eq:ME_Redfield_final} we compare our results with numerically exact solution as obtained using HEOM~\cite{Huang2023}. It should be noted that in the HEOM~calculation, the particle super selection rule is not forced, contrary what was done to derive Eq.~\eqref{Eq:ME_Redfield_final}, allowing for dot superposition states. Due to the computational costs of the exact solutions, in this section we restrict ourselves to the low temperature case, where the \resonator~state can be well approximated by a truncation to $\leq 10$ levels. 

\begin{figure}[t]
    \centering
    \includegraphics[width=\linewidth]{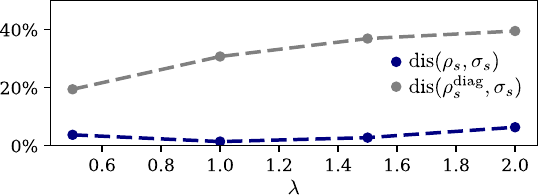}
    \caption{
    \textbf{Benchmark of \NESS~$\rho_s$ compared to $\sigma_s.$}
    Trace distance between the \NESS~using the Redfield equation derived in this work $\rho_s$ and the exact solution computed using HEOM~$\sigma_s$ as a function of \qd+\resonator~coupling $\gw$ in blue. For comparison, in gray we exhibit the trace distance between a classical analogous of our solution $\rho_s^\text{diag}$ with $\sigma_s$.
    Parameters: $\Gamma_L/2\pi=\Gamma_R/2\pi=0.1$ GHz, $\omega/2\pi$= 1 GHz, $T_L=T_R=2$ GHz (corresponds to $15.28$ mK), $\gamma_L=-\gamma_R= 2.5$ GHz, $\delta_L=\delta_R=2$ GHz, $\mu_L=-\mu_R=2.5$ GHz, and $\tilde{\mu}=0$ GHz.}
    \label{fig:Figure3}
\end{figure}

First, we will quantify how close our derived \NESS~($\rho_s$) is to the exact solution ($\sigma_s$). As a measure we use the trace distance~\cite{Nielsen2010}
\begin{equation}\label{Eq:trace_distance}
    \text{dis}(\rho_s,\sigma_s) = \frac{1}{2} \Tr\Big( \sqrt{(\rho_s-\sigma_s)^\dagger(\rho_s-\sigma_s)}\Big).
\end{equation}
In Fig.~\ref{fig:Figure3} we plot in blue this distance as a function of increasing \qd+\resonator~coupling. We consider the resonance frequency of the \resonator~to be $\omega/2\pi=1$ GHz and the tunneling rates an order of magnitude smaller, i.e. $\Gamma_L/2\pi=\Gamma_R/2\pi=0.1$ GHz, well within the slow tunneling regime achievable with the equation derived in this work. We observe that for all couplings the distance between the exact HEOM solution and the derived Redfield does not surpass 10\% discrepancy. For further comparison, we also compute the distance between the exact solution and a `classical' analogous of $\rho_s$, $\rho_s^\text{diag}$, where energetic coherences in the~\resonator~are ignored in the master equation (see Appendix~\ref{Appendix:Classical_analogues}). In Fig.~\ref{fig:Figure3} we see that the distance between $\sigma_s$ and $\rho_s^\text{diag}$ (gray) is significantly higher than the distance with the Redfield steady-state $\rho_s$. This showcases that indeed in the regime considered here, the presence of energetic coherence in the \resonator~are required to reliably compute the NEMS steady-state.

\begin{figure}[t]
    \centering
    \includegraphics[width=\linewidth]{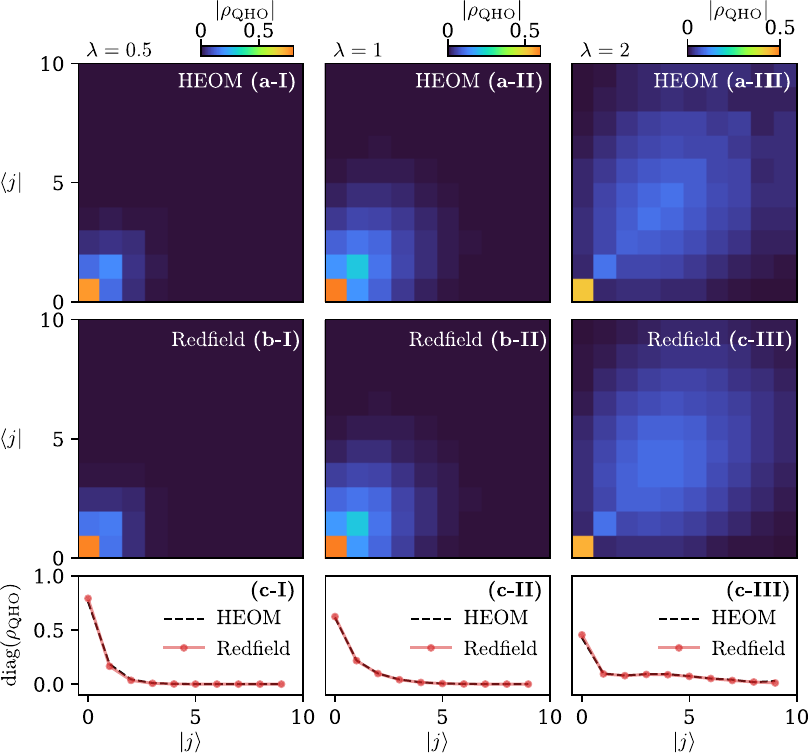}
    \caption{\textbf{Non-equilibrium steady state elements of the \resonator~density matrix $\rho_\text{\resonator}$.}
    The upper rows are solved using HEOM~for increasing \qd+\resonator~couplings $\gw=0.5$ \textbf{(a-I)},  $\gw=1$ \textbf{(a-II)}, and $\gw=1.5$ \textbf{(a-III)}.
    The lower rows are the corresponding elements obtained using the Redfield equation derived in this work (see Eq.~\eqref{Eq:ME_Redfield_final}). 
    Color-map shows the magnitude of the element of the reduced density matrix of the \resonator. 
    \textbf{(c-I)}-\textbf{(c-III)} Populations of the \resonator~state corresponding to exact solution in dashed black and obtained via Eq.~\eqref{Eq:ME_Redfield_final} in red.
    Parameters: $\Gamma_L/2\pi=\Gamma_R/2\pi=0.1$ GHz, $\omega/2\pi$= 1 GHz, $T_L=T_R=2$ GHz (corresponds to $15.28$ mK), $\gamma_L=-\gamma_R= 2.5$ GHz, $\delta_L=\delta_R=2$ GHz, $\mu_L=-\mu_R=2.5$ GHz, and $\tilde{\mu}=0$ GHz.}
    \label{fig:Figure4}
\end{figure}

Next, we take a closer look at the reduced state of the \resonator, $\rho_\text{\resonator}=\Tr_\text{\qd}(\rho_s)$. In Sec.~\ref{Sec:Current} we will look at the electronic particle current, which will give us a greater insight on the state of the \qd. In Fig.~\ref{fig:Figure4} we plot $\rho_\text{\resonator}$ in the \resonator-Fock basis defined in Eq.~\eqref{eq:Energy_system}. In Figs.~\subfigref{fig:Figure4}{a-I}-\subfigref{fig:Figure4}{a-III} we showcase the numerically exact solutions provided by HEOM~while in Figs.~\subfigref{fig:Figure4}{b-I}-\subfigref{fig:Figure4}{b-III} the ones obtained using Eq.~\eqref{Eq:ME_Redfield_final} for increasing \qd+\resonator~coupling, namely \textbf{(I)} $\gw=0.5$, \textbf{(II)} $\gw=1$ and \textbf{(III)} $\gw=1.5$. Overall, we observe an excellent agreement between the solutions irrespective of the coupling strength. This is due to our use of the polaron transformation, which allows us to obtain dependable results even in the ultrastrong coupling regime ($\gw\gg1$) without having to make use of perturbation theory. Moreover, we observe that the state of the \resonator~exhibits coherences which are enhanced for larger coupling $\gw$, as was pointed out in Ref.~\cite{Piovano2011}. In
Figs.~\subfigref{fig:Figure4}{c-I}-\subfigref{fig:Figure4}{c-III} we compare the energy basis occupations of $\rho_\text{\resonator}$. For all couplings we again observe good agreement between our master equation and HEOM. We also note that, while the ground state of the QHO is the most likely state, at high couplings we observe a non-monotonous distribution for the occupations, which is a signature of non-thermal steady states of the QHO~\cite{Culhane2022,Sevitz2025}.
\begin{figure*}[t]
    \centering
    \includegraphics[width=\linewidth]{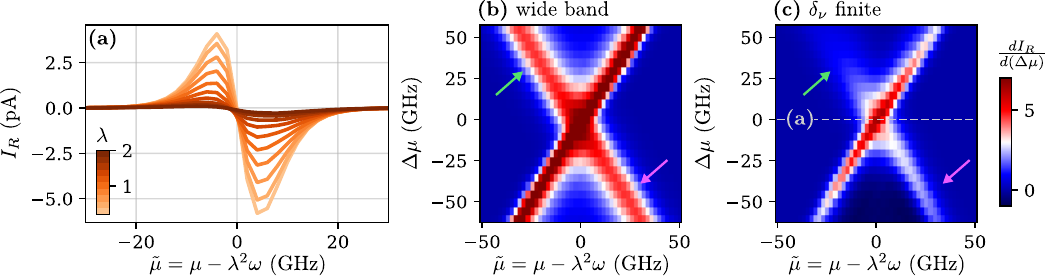}
    \caption{
    \textbf{Particle current obtained from the microscopic model.}
    \textbf{(a)} Particle current corresponding to the right reservoir, $\IR$ as function of the dot's normalized energy $\tilde{\mu}$ for increasing dot-resonator coupling strength $\gw$. Here there is no chemical potential imbalance $\Delta\mu= \mu_L-\mu_R=0$ GHz and the electronic transport is solely due to the temperature imbalance among the reservoirs. 
    Stability diagram d$\IR$/$d\Delta\mu$ as a function of $\tilde{\mu}$ and $\Delta\mu$ for a \qd+\resonator~coupling of $\gw=1$ under wide-band in panel \textbf{(b)} and for finite spectral density in panel \textbf{(c)}. Both panels share the same color-map located to the right. Green and purple arrows indicate the asymmetric current suppression with respect to $\Delta\mu=0$ GHz present in the finite spectral density case.
    Parameters: $\Gamma_L/2\pi=\Gamma_R/2\pi=0.1$ GHz, $\omega/2\pi$= 1 GHz, $T_L= 2$ GHz (corresponds to $15.28$ mK), $T_R=4.5$ ($34.4$ mK) , $\gamma_L=-\gamma_R= -10$ GHz, $\delta_L=20$ GHz, and $\delta_R=10$ GHz.
    }
    \label{fig:Figure5}
\end{figure*}


\subsection{Note on dynamics}\label{Sec:Note_dynamics}

In this work, the main focus is on the \NESS~since, for many experimental platforms, this is the only regime that can directly be probed. However, Eq.~\eqref{Eq:ME_Redfield_final} is also able to describe the transient dynamics of the NEMS.
At finite time, some care is needed with the choice of parameters to ensure the validity of the Markov approximation. In practice, this means that the bath correlation times must be faster than the characteristic timescales of the system dynamics~\cite{Potts2024} \footnote{This was not a problem previously when we studied the \NESS~given that this condition is reduced to making sure that the bath correlation function decay at some time scale~\cite{Strasberg2022Book}, see Appendix~\ref{Appendix:Bath_convergance} for details.}. At this point, we further note that the Redfield equation does not yield a completely positive trace-preserving (CPTP) map and can therefore result in non-physical states~\cite{BreuerPetruccione2002,Hartmann2020,Montoya2015}. For this reason, it is customary to carry out a secular proximation and arrive at a master equation in the well-known GKLS form, which guarantees a CPTP evolution~\cite{BreuerPetruccione2002}. In Appendix~\ref{Appendix:LindblatForm} we carry out the secular approximation valid for the regime
\begin{equation}\label{Eq:secular_aprox_bound}
    \Gamma_\nu \ll 2 \max\{\tilde{\mu},\omega\} \quad \forall \, \nu=L,R,
\end{equation}
see also Appendix~\ref{Appendix:Secular_aprox_cond}.
We note that the GKLS equation obtained in (Eq.~\eqref{Eq_app:Lindblat_projected}) is valid for any spectral density. However, for spectral densities other than Lorentzian, the the condition for the validity of the secular approximation, Eq.~\eqref{Eq:secular_aprox_bound}, will change.


\section{The electronic particle current}\label{Sec:Current}

We now move to look at the particle current in the system, which is the most easily accessed quantity experimentally. Formally, the average current through the reservoir $\nu$ is defined as $\I = e\, \partial_t\langle \hat{N}_\nu \rangle$, where $e$ is the electron charge, $\hat{N}_\nu=\sum_k \cc \c$ the reservoir number operator and $\langle \bullet \rangle = \Tr(\bullet \rho^\text{NESS})$~\cite{Schaller2014Book, Leijnse2010, Potts2024, Josefsson2018, Wachtler2019}. The sign convention is such that the currents are positive if the particle flows towards reservoir $\nu$, and negative otherwise. As shown in Appendix~\ref{Appendix:I_computation}, from the Redfield equation \eqref{Eq:ME_Redfield_final} we find
\begin{align}\label{Eq:particle_current}
    \I &= - e\, \sum_{qkl} \left( \Rnn{0}{1}{q}{q}{k}{l} \Pbra{k} \trho_s^{0}\Pket{l} - \Rnn{1}{0}{q}{q}{k}{l} \Pbra{k} \trho_s^{1}\Pket{l}  \right).
\end{align}
It is straightforward to show that this expression reduces to the conventional \qd~electronic current for the un-coupled dot-resonator case ($\lambda=0$)~\cite{Potts2024}. In what follows, we show that Eq.~\eqref{Eq:particle_current} captures the expected behavior of electronic transport and compare with previous experimental results. 

As we are interested in the \NESS~where the total particle current is conserved, i.e. $\IL+\IR=0$, it is sufficient to consider the $\nu=R$ case. In Fig.~\subfigref{fig:Figure5}{a} we plot $\IR$ as a function of the dot's re-normalized energy $\tilde{\mu}$ for increasing \qd+\resonator~coupling $\gw$. We take a temperature imbalance with $T_L>T_R$ while setting $\mu_L=\mu_R$. We see that $\IR$ changes sign for negative values of $\tilde{\mu}$, a well known signature of thermally driven currents in \qd~systems~\cite{Dorsch2021, Josefsson2018}. Furthermore, we note that for larger $\gw$, the particle current $|\IR|$ diminishes, which is consistent with well-known Franck–Condon blockade regime~\cite{Koch2005}. Therefore, the particle currents obtained here are able to recover expected effects in previously studied regimes.

Finally, we compare the current obtained with and without energy-dependent tunneling rates. In Fig.~\subfigref{fig:Figure5}{b} we plot the charge stability diagram as a function of $\tilde{\mu}$ and $\Delta\mu$ for $\gw=1$ under the wide-band approximation with $\Gamma_R = \Gamma_L$. The plot of the differential conductance shows the well known shape of Coulomb diamonds. If we now instead take a spectral density with finite width, as shown in Fig.~\subfigref{fig:Figure5}{c}, there is a clear suppression of the conductance at the edges of the Coulomb diamond. Moreover, this suppression is asymmetric with respect to the zero bias point, as indicated by the green and purple arrows. These features are consistent with previous experiments, such as in Fig. 2(a) of Ref.~\cite{Meerwaldt2012}.


\section{Conclusions}

In this article we have developed a theoretical model to describe the dynamics of NEMS~in the quantum domain, with slow electronic transport, and keeping energy dependence of tunneling rates. Specifically, we have focused our analysis on the single electron tunneling regime by deriving a second order (Redfield) master equation. Although our approach is able to capture transient dynamics, we have focused our study on the \NESS~given that it is the most readily accessible time scale experimentally. To benchmark the obtained \NESS~we compared  with the exact result obtained by the HEOM. Overall our solutions matches the ones obtained with HEOM for all \qd-\resonator~coupling strength. Moreover, we derived an expression for the steady-state particle current, which is able reproduce well-known behavior such as thermal currents and Franck–Condon blockade and further showcases features of the presence of energy-dependent tunneling barriers.

Our work provides the essential tools to explore applications of \qd~in NEMS devices in a regime where previously an applicable model was missing. Potential areas of interest for this regime are quantum thermodynamics and quantum information. For example, the model developed here allows for the first time to explore the appearance of self-oscillations in the slow transport regime, and the application of the nanomechanical resonator as a work load or as a battery.


\begin{acknowledgments}
The authors would like to acknowledge K. Hovhannisyan for the fruitful conversations and valuable feedback. S.S. would also like to thank J. Tabanera-Bravo, M. Aguilar, and F. Hartmann for the inspiring conversations. S.S. and J.A. acknowledge the support by Deutsche Forschungsgemeinschaft (DFG 384846402). J.A. and F.C. acknowledge support from EPSRC (EP\/R045577\/1). J.A. thanks the Royal Society for support. Computation of the \NESS~were performed with computing resources provided by ZIM, Universität Potsdam. 
\end{acknowledgments}

\bibliography{bibliography}
\appendix
\onecolumngrid

\section{Derivation of master equation}\label{Appendix:ME}

In this appendix we give all the details to derive the master equation that describes the system dynamics of the system density matrix  $\trho_s$ associated with the \qd+\resonator~system under the polaron picture stated in Eq.~\eqref{eq:H_s_tilde} of the main text.

We start the derivation by considering the evolution of the total density matrix including the two reservoirs associated to the polaron transformed Hamiltonian is 
\begin{equation}\label{Eq_app:Full_evolution_under_polaron}
    \partial_t \trho(t)= -i [\tilde{H},\trho(t)] .
\end{equation}
At this point we follow the formalism detailed in Ref.~\cite{Mitra2004}, for self-consistency we will go over the key points here. Here they define two density matrices: the uncorrelated and the correlated as
\begin{align}\label{Eq_app:rho_un_correlated}
    \trho_{\otimes}(t) &= \trho_{s}(t)\otimes \rhoR \\ \label{Eq_app:rho_correlated}
    \trho_{\circ}(t)&= \trho(t) - \trho_{\otimes}(t)
\end{align}
respectively. In Eq.~\eqref{Eq_app:rho_un_correlated} we made use of the density matrix of the fermionic reservoirs at thermal equilibrium 
\begin{equation}\label{eq_app:thermal_rhoR}
    \rhoR = \bigotimes_{\nu} \frac{1}{Z_\nu}\exp\left\{- \beta_\nu\left(\sum_k (\epsilon_{\nu k}-\mu_\nu) \cc\c\right)\right\}
\end{equation}
where $\epsilon_{\nu k}$ is the free energy of a spin-less electron with wave vector $k$, $\mu_\nu$, $\beta_\nu=1/T_\nu$ the inverse temperature and $Z_\nu$ the normalization factor for the reservoir $\nu$. The dynamics of $\trho_\otimes$ yields the dynamics of the system (\qd~+\resonator) while the dynamics of $\trho_\circ$ is going to be used when calculating the transport properties in Sec.~\ref{Sec:Current}.

At this point it is convenient to go to the interaction picture defined as
\begin{equation}\label{Eq_app:Interaction}
    \Int{(\cdot)}=e^{i(\tilde{H}-\tVr)t}(\cdot)e^{-i(\tilde{H}-\tVr)t},
\end{equation}
leaving Eq.~\eqref{Eq_app:Full_evolution_under_polaron} as 
\begin{equation}\label{Eq_app:Full_evolution_under_polaron_I}
    \partial_t{\trho}_I(t)=-i\left[\Int{\tilde{V}}(t),\trho_I(t)\right],
\end{equation}
where $\Int{\tilde{V}}$ is the tunneling coupling defined in Eq.~\eqref{eq:V_r_tilde} from the main text.
Making use of Eq.~\eqref{Eq_app:Full_evolution_under_polaron_I} we obtain the dynamics of both $\trho_\otimes$ and $\trho_\circ$ as
\begin{align}\label{Eq_app:rho_uncorrelated_ME}
    \partial_t\Int{(\trho_\otimes)}(t) &= \partial_t\Int{(\trho_s)}(t) \otimes \rhoR  = \Tr_\R\left(\partial_t\Int{(\trho)}(t) \right) \otimes \rhoR  = -i\,\Tr_\R\Big([\tVrI(t),\Int{(\trho_\circ)}(t)\Big) \otimes \rhoR \\  \label{Eq_app:rho_correlated_ME}
    \partial_t \Int{(\trho_\circ)}(t) &= \partial_t\Int{\trho}-\partial_t \Int{(\trho_\otimes)} =  -i\left[\Int{\tilde{V}}(t),\trho_I(t)\right] +i\, \Tr_\R \Big(\left[\Int{\tilde{V}}(t),\Int{(\trho_\circ)}(t)\right] \Big) \otimes \rhoR
\end{align}
respectively. Now we take the widely used system-reservoir un-correlated initial conditions~\cite{Timm2008,Mitra2004,Piovano2011}
\begin{equation}\label{Eq_app:initial_state}
    \trho_{s}(t_0=-\infty)\otimes \rhoR.
\end{equation}
For this initial boundary condition, Eq.~\eqref{Eq_app:rho_correlated_ME} can be formally solved as 
\begin{equation}\label{Eq_app:tho_circ_sol}
    \Int{(\trho_\circ)} (t)= -i \int_{-\infty}^\infty dt'\, K_I (t,t') [\tVrI(t'),\Int{(\trho_\otimes)}(t')]
\end{equation}
where $K_I(t,t')$ is a kernel function solution to the equation~\cite{Mitra2004}
\begin{equation}\label{Eq_app:tho_circ_sol_weak_before}
    \frac{d}{d t} \left( K_{I}(t, t') \bullet \right)+i[\tVrI, K_{I}\left(t, t' \right) \bullet] -i \Tr_\R [\tVrI, K_{I}(t, t') \bullet] \otimes \rhoR =\delta(t-t'),
\end{equation}
where $\bullet$ is an arbitrary operator. As mentioned in the main text, we restrict our analysis to the single electron tunneling regime $\Gamma_\nu \ll T_\nu$ also known as the \textit{high temperature limit} where the kernel can be approximated as
$K_I(t,t')\rightarrow \theta(t-t')$, with $\theta$ the Heaviside step function. 
This results in 
\begin{equation}\label{Eq_app:tho_circ_sol_weak}
    \Int{(\trho_\circ)} (t)= -i \int_{-\infty}^t dt'\, [\tVrI(t'),\Int{(\trho_\otimes)}(t')].
\end{equation}
We refer the reader to Ref.~\cite{Mitra2004} for more details. This \textit{high temperature limit} may sound a bit vague at first where the reader might question the consideration from Eq.~\eqref{Eq_app:tho_circ_sol_weak_before} to Eq.~\eqref{Eq_app:tho_circ_sol_weak}. For the transport community, this approximation is understood as a second order perturbation in the \qd-reservoirs coupling (termed as \textit{weak coupling} for the open quantum systems community) to describe the system dynamics~\cite{Timm2008}. This is what we will obtain at the end of this computation ultimately justifying the derivation from Eq.~\eqref{Eq_app:tho_circ_sol_weak_before} to Eq.~\eqref{Eq_app:tho_circ_sol_weak}, see Eq.~\eqref{Eq_app:ME-local}. Future work will focus on a concrete derivation to relate high temperature regime from the transport community with \textit{weak coupling} form the quantum open system community.

Resuming our calculation, we replace Eq.~\eqref{Eq_app:tho_circ_sol_weak} into Eq.~\eqref{Eq_app:rho_uncorrelated_ME} that yields 
\begin{equation}\label{Eq_app:rho_uncorrelated_ME2}
    \partial_t\Int{(\trho_\otimes)}(t) = -\int_{-\infty}^t dt' \, \Tr_\R\Big(\left[\tVrI(t),\left[\tVrI(t'),\Int{(\trho_\otimes)}(t')\right]\right]\Big) \otimes \rhoR 
\end{equation}
As we are searching for the dynamics of the system $\trho_s$, we trace out the reservoirs from Eq.~\eqref{Eq_app:rho_uncorrelated_ME2} and take following change of variables $t'=t-s$ resulting in the following master equation that is non-local in time
\begin{align}\label{Eq_app:ME-non-local}
    \partial_t \Int{(\trho_s)}(t) = - \int_{0}^{+\infty} ds \, \Tr_\R\Big(\left[\tVrI(t),\left[\tVrI(t-s),\Int{(\trho_s)}(t-s)\otimes\rhoR \right]\right]\Big).
\end{align}
We made use of the definition of $\trho_{\otimes I}(t')$ from Eq.~\eqref{Eq_app:rho_un_correlated}. The deduction done here offers an alternative to the so-called Born approximation, i.e. $\trho(t)= \trho_s(t) \otimes \rhoR \, \forall t$, to preserve the correlations among system and reservoir justifying the study of the transport properties in the \NESS. We note that in Ref.~\cite{Potts2024} a similar approach is taken that using the Nakajima-Zwanzig superoperators that also lead to Eq.~\eqref{Eq_app:ME-non-local}.

The next step consist in making Eq.~\eqref{Eq_app:ME-non-local} local in time, that is the state at time $t$ is only determined by the state at the same time. This is justified if the bath correlation times $\tau_\R$ are very short. Alternatively, this is always justified in the \textit{long time scale} regime we are interested in this work~\cite{Strasberg2022Book}. Thus this requirement is reduced to checking that the bath correlation functions effectively decay at some time. 
With this in mind, Eq.~\eqref{Eq_app:ME-non-local} reduces to the Redfield equation (or Born-Markov equation) in the interaction picture
\begin{align}\label{Eq_app:ME-local}
    \partial_t \Int{(\trho_s)}(t) = - \int_{0}^{+\infty} ds \, \Tr_\R\Big(\left[\tVrI(t),\left[\tVrI(t-s),\Int{(\trho_s)}(t)\otimes\rhoR \right]\right]\Big).
\end{align}

We can go a step further and write the tunneling Hamiltonian defined in Eq.~\eqref{eq:V_r_tilde} (in the Schrödinger picture) from the main text in the following form 
\begin{equation}\label{eq:Coupling_expression}
    \tVr = \sum_{\nu=L,R} \left(\d e^{-\gw (\b^\dagger-\b)} \sum_{k} \left( \t \cc \right) +\dd e^{+\gw (\b^\dagger-\b)} \sum_{k} \left( \tt \c   \right)  \right) 
    = \sum_{\nu=L,R} \left( \S_{0} \B_{\nu 0} + \S_{1} \B_{\nu 1} \right),
\end{equation}
such that the $\S$ operators belong to the system as $\S_{0} = \d \hat{D}(-\gw)$ and $\S_{1} = \dd \hat{D}(+\gw)$, while $\B$ belong to the reservoirs 
\begin{align}\label{Eq_app:Bath_operators_int}
    \Int{(\B_{\nu 0})}(t) = \sum_k \t e^{i\eps t} \cc   &&
    \Int{(\B_{\nu 1})}(t) = \sum_k \tt e^{-i\eps t} \c. 
\end{align}
We remind the reader the tilde denotes that we are working under the polaron picture. At this point we remind the reader that we define the bath operators under the Klein transformation such that $[\c,\dd]=[\cc,\d]=0$. Replacing this into Eq.~\eqref{Eq_app:ME-local} yields~\cite{Potts2024,Strasberg2022Book}
\begin{align}\nonumber
    \partial_t \Int{(\trho_s)}(t)= \sum_\nu \sum_{\substack{q=0,1 \\ q'= 0,1}} \int^\infty_0 & ds\Bigg( C^\nu_{qq'}(s) \left( \Int{(\S_{q'})}(t-s) \, \Int{(\trho_s)}(t) \, \Int{(\S_{q})}^\dagger(t) - \Int{(\S_{q})}^\dagger(t) \, \Int{(\S_{q'})}(t-s) ', \Int{(\trho_s)}(t)\right)  \\ \label{Eq_app:Redfield_general_Interaction}
    & + C^\nu_{qq'}(-s) \left( \Int{(\S)_{q'}}(t)\, \Int{(\trho_s)}(t) \, \Int{(\S_{q})}^\dagger(t-s) - \Int{(\trho_s)}(t) \, \Int{(\S_{q})}^\dagger(t-s) \, \Int{(\S_{q'})}(t)\right) \Bigg)
\end{align}
where 
\begin{equation}\label{eq:Bath_correlation_func}
    C_{q q'}^\nu (s)=\Tr_\R\left( \Int{(\B_{\nu q})}^\dagger(s) \Int{(\B_{\nu q'})}(0) \rhoR\right) 
\end{equation}
are the bath correlation functions. Making use of Eq.~\eqref{Eq_app:Bath_operators_int} it is straightforward to see that~\cite{Potts2024} 
\begin{align}\label{Eq_app:C10}
    C^\nu_{10}(s) &= \Tr_\R\left( \Int{(\B_{\nu 1})}^\dagger(s) \Int{(\B_{\nu 0})}(0) \rhoR \right) = 0\\ 
     \label{Eq_app:C00}
    C^\nu_{00}(s) &= \Tr_\R\left( \Int{(\B_{\nu 0})}^\dagger(s) \Int{(\B_{\nu 0})}(0) \rhoR \right) = \frac{1}{2\pi}\int_{-\infty}^\infty d\omega' \, e^{-i \omega' s} \G(\omega') [1-\f(\omega' )]\\ \label{Eq_app:C11}
     C^\nu_{11}(s) & = \Tr_\R\left( \Int{(\B_{\nu 1})}^\dagger(s) \Int{(\B_{\nu 1})}(0) \rhoR \right) = \frac{1}{2\pi} \int_{-\infty}^\infty d\omega' \, e^{+i\omega' s} \G(\omega') \f(\omega') 
    \\ \label{Eq_app:C01}
     C^\nu_{01}(s) &= \Tr_\R\left( \Int{(B_{\nu 0})}^\dagger(s) \Int{(B_{\nu 1})}(0) \rhoR \right) = 0
\end{align} 
where $\G(\omega)$ is the spectral density of lead $\nu$ related to the tunneling coupling constants $\t$ as $\G(\omega)=2\pi \sum_k |\t|^2\delta(\omega-\eps)$ and  $\f(\omega)=1/(\exp(\beta_\nu(\omega-\mu_\nu))+1)$ is the Fermi function, with $\mu_\nu$ their respective chemical potentials and $\beta_\nu=1/T_\nu$ inverse temperatures. Plugging Eq.~\eqref{Eq_app:C10}-\eqref{Eq_app:C01} into Eq.~\eqref{Eq_app:Redfield_general_Interaction} gives
\begin{align}\nonumber 
    \partial_t \Int{(\trho_s)}(t)= \sum_{\nu} \sum_{q=0,1} \int^\infty_0 ds\Bigg( &C^\nu_{qq}(s) \left( \Int{(\S_q)}(t-s) \, \Int{(\trho_s)}(t) \, \Int{(\S_q)}^\dagger(t)- \Int{(\S^\dagger_q)}(t) \,  \Int{(\S_q)}(t-s) \, \Int{(\trho_s)}(t)\right)  \\ \label{Eq_app:Redfield_simple_Interaction}
    &+ C^\nu_{qq}(-s) \left(\Int{(\S_q)}(t) \, \Int{(\trho_{s})} (t) \, \Int{(\S_q)}^\dagger(t-s) - \Int{(\trho_s)}(t) \, \Int{(\S_q)}^\dagger(t-s) \, \Int{(\S_q)}(t)\right) \Bigg)
\end{align}

\subsection{Projection of Redfield}\label{Appendix:Redfield}

To compute the non-equilibrium steady state defined in Eq.~\eqref{eq:NESS_def} of the main text it is convenient to project the density matrix into the energy eigenbasis [see Eq.~\eqref{eq:Energy_system} of the main text]. First we return to the Schrödinger picture as
\begin{align}\nonumber 
    \partial_t \trho_s(t) =  -i [\tHs,\trho_s(t)]
    +\sum_{\nu} \sum_{q=0,1} \int^{+\infty}_0 ds\Bigg( &C^\nu_{qq}(s) \left(e^{-i \tHs s} \S_q e^{+i \tHs s} \trho_s(t) \S_q^\dagger- \S^\dagger_q e^{-i \tHs s} \S_q e^{+i \tHs s} \trho_s(t)\right) \\ \label{Eq_app:Redfield_simple_Schrodinger}
    &+C^\nu_{qq}(-s) \left( \S_q  \trho_{s}(t) e^{-i \tHs s} \S_q^\dagger e^{+i \tHs s} - \trho_s(t) e^{-i \tHs s} \S_q^\dagger e^{+i \tHs s} \S_q\right) \Bigg)
\end{align}
We remind the reader that the tilde denotes that the density matrix is defined under the polaron transformation. Now we project into the basis defined in Eq.~\eqref{eq:Energy_system} of the main text resulting in
\begin{align}\nonumber 
    \Fbra{j}{n}\partial_t \trho_s(t) \Fket{m}{n} =  -i (\E{j}{n}-\E{m}{n}) &\Fbra{j}{n} \trho_s (t)\Fket{m}{n} \\ \nonumber
    +\sum_{\nu=L,R} \sum_{q=0,1} \sum_{i,l=0}^N \sum_{\tau,\sigma=0}^1 \int^\infty_0 ds\Bigg(
    &C^\nu_{qq}(s) \Big(\delta_{\tau,\sigma} e^{-i (\E{j}{n}-\E{i}{\tau})s}   \Fbra{j}{n} \S_q \Fket{i}{\tau}\Fbra{i}{\tau}  \trho_s(t) \Fket{l}{\sigma}  \Fbra{l}{\sigma} \S_q^\dagger \Fket{m}{n} \\ \nonumber
    &- \delta_{\sigma,n} e^{-i (\E{i}{\tau}-\E{l}{\sigma}) s} \Fbra{j}{n} \S^\dagger_q  \Fket{i}{\tau}\Fbra{i}{\tau} \S_q \Fket{l}{\sigma}\Fbra{l}{\sigma} \trho_s(t)\Fket{m}{n}  \Big)  \\ \nonumber
    &+C^\nu_{qq}(-s) \Big( \delta_{\tau,\sigma} e^{-i (\E{l}{\sigma}-\E{m}{n}) s} \Fbra{j}{n}\S_q \Fket{i}{\tau}\Fbra{i}{\tau} \trho_{s}(t) \Fket{l}{\sigma} \Fbra{l}{\sigma}  \S_q^\dagger\Fket{m}{n} \\
    &- \delta_{n,\tau}\Fbra{i}{\tau} e^{-i (\E{i}{\tau}-\E{l}{\sigma}) s} \Fbra{j}{n} \trho_s(t)  \Fket{i}{\tau}  \S_q^\dagger \Fket{l}{\sigma}\Fbra{l}{\sigma} \S_q \Fket{m}{n}\Big) \Bigg) 
\end{align}
where we have included some identities that yield the sum over indices $i,\tau$ and $l,\sigma$ and $\E{j}{n}$ is defined in Eq.~\eqref{eq:Energy_system} of the main text. The $\delta$'s arise from the fact that we are only considering the diagonal elements of the dot degree due to particle super selection rule~\cite{Potts2024, Bartlett2007}. In contrast, the mechanical part this is not always diagonal, as was shown in Ref.~\cite{Piovano2011}. Specifically it is strongly dependent on the \qd+\resonator~coupling $\gw$ and the temperature of the reservoirs $T_\nu$. For a simplified notation we take $\Fbra{j}{n}\trho_s(t) \Fket{m}{n}=\Pbra{j}\trho^n_s(t) \Pket{m}$
\begin{align}\nonumber 
    \Pbra{j}\partial_t \trho^n_s(t) \Pket{m} =  -i (\E{j}{n}-\E{m}{n}) &\Pbra{j} \trho^n_s (t) \Pket{m} \\ \nonumber
    +\sum_{\nu=L,R} \sum_{q=0,1} \sum_{i,l=0}^N \sum_{\sigma=0}^1 \int^{+\infty}_0 \mathrm{d}s \, \Bigg(
    &C^\nu_{qq}(s) \Big(e^{-i (\E{j}{n}-\E{i}{\sigma})s}   \Fbra{j}{n} \S_q \Fket{i}{\sigma}  \Fbra{l}{\sigma} \S_q^\dagger \Fket{m}{n} \Pbra{i}\trho^\sigma_s(t)\Pket{l} \\ \nonumber
    &-  e^{-i (\E{i}{\sigma}-\E{l}{n}) s} \Fbra{j}{n} \S_q^\dagger  \Fket{i}{\sigma}\Fbra{i}{\sigma} \S_q \Fket{l}{n}\Pbra{l} \trho^n_s(t)\Pket{m}  \Big)  \\ \nonumber
    &+C^\nu_{qq}(-s) \Big( e^{-i (\E{l}{\sigma}-\E{m}{n}) s} \Fbra{j}{n}\S_q \Fket{i}{\sigma} \Fbra{l}{\sigma}  \S_q^\dagger\Fket{m}{n}\Pbra{i}\trho_s^\sigma(t)\Pket{l}\\ 
    &-  e^{-i (\E{i}{n}-\E{l}{\sigma}) s} \Fbra{i}{n} \S_q^\dagger \Fket{l}{\sigma} \Fbra{l}{\sigma} \S_q \Fket{m}{n} \Pbra{j}\trho^n_s(t)\Pket{i} \Big) \Bigg) .
\end{align}
To simplify notation we use the explicit expression of $\S_0^\dagger=\S_1=\dd \D(\gw)$ and
$\S_1^\dagger=\S_0=\d \D(-\gw)$ we define
\begin{align}\label{Eq_app:S_0_elements}
    &\Fbra{a_1}{n_1} \S_{0} \Fket{a_2}{n_2} = \Fbra{a_1}{n_1} \S_{1}^\dagger \Fket{a_2}{n_2} =  \Fbra{a_1}{n_1} \d e^{-\gw(\bb-\b)}  \Fket{a_2}{n_2} = \delta_{n_1,0}\delta_{n_2,1} \DD_{a_1, a_2}\\ \label{Eq_app:S_1_elements}
    &\Fbra{a_1}{n_1} \S_{1} \Fket{a_2}{n_2} = \Fbra{a_1}{n_1} \S_{0}^\dagger \Fket{a_2}{n_2} = \Fbra{a_1}{n_1} \dd e^{+\gw(\bb-\b)} \Fket{a_2}{n_2} = \delta_{n_1,1}\delta_{n_2,0} \D_{a_1,a_2}
\end{align}
here we considered that the electronic states are $n_1,n_2=\{0,1\}$ and the $\D_{a_1a_2},\DD_{a_1a_2}$ is the element $a_1,a_2$ of the displacement operator.
\begin{align}\nonumber 
    \Pbra{j}\partial_t \trho^n_s(t) \Pket{m} &=  -i (\E{j}{n}-\E{m}{n}) \Pbra{j} \trho^n_s (t) \Pket{m} 
    +\sum_{\nu} \sum_{i,l} \sum_{\sigma}  \\ \nonumber \int^{+\infty}_0 \mathrm{d}s \, \Bigg( &
    C^\nu_{00}(s) \Big(e^{-i (\E{j}{n}-\E{i}{\sigma})s}   
    \SZeroSOneD{j}{n}{i}{\sigma} \D_{l,m} \Pbra{i}\trho^\sigma_s(t)\Pket{l} 
    -  e^{-i (\E{i}{\sigma}-\E{l}{n}) s} 
    \SOneSZeroD{j}{n}{i}{\sigma} \DD_{i,l} \Pbra{l}\trho^n_s(t)\Pket{m}  \Big)  \\ \nonumber
    &+C^\nu_{00}(-s) \Big( e^{-i (\E{l}{\sigma}-\E{m}{n}) s} 
    \SZeroSOneD{j}{n}{i}{\sigma} \D_{l,m} \Pbra{i}\trho^\sigma_{s}(t)\Pket{l}
    -  e^{-i (\E{i}{n}-\E{l}{\sigma}) s} 
    \SOneSZeroD{i}{n}{l}{\sigma} \DD_{l,m} \Pbra{j}\trho^n_s(t)\Pket{i} \Big)  \\ \nonumber
    &+C^\nu_{11}(s) \Big(e^{-i (\E{j}{n}-\E{i}{\sigma})s}   
    \SOneSZeroD{j}{n}{i}{\sigma} \DD_{l,m} \Pbra{i}\trho^\sigma_s(t)\Pket{l} 
    -  e^{-i (\E{i}{\sigma}-\E{l}{n}) s} 
    \SZeroSOneD{j}{n}{i}{\sigma} \D_{i,l} \Pbra{l} \trho^n_s(t)\Pket{m}  \Big)  \\ 
    &+C^\nu_{11}(-s) \Big( e^{-i (\E{l}{\sigma}-\E{m}{n}) s} 
    \SOneSZeroD{j}{n}{i}{\sigma} \DD_{l,m} \Pbra{i}\trho^\sigma_{s}(t)\Pket{l}
    -  e^{-i (\E{i}{n}-\E{l}{\sigma}) s} 
    \SZeroSOneD{i}{n}{l}{\sigma} \D_{l,m} \Pbra{j}\trho^n_s(t)\Pket{i} \Big) \Bigg) 
\end{align}
Next we define $\e{k}{l}=\tilde{\mu}-\omega(k-l)$ that results in 
\begin{align}\nonumber 
    \Pbra{j}\partial_t \trho^n_s(t) \Pket{m} &=  -i (\E{j}{n}-\E{m}{n}) \Pbra{j} \trho^n_s (t) \Pket{m} \\ \nonumber
    +\sum_{\nu}  \sum_{i,l}  \int^{+\infty}_0 \mathrm{d}s \, \Bigg(
    &C^\nu_{00}(s) \Big(\delta_{n,0} e^{-i (-\e{j}{i})s} \DD_{j,i}\D_{l,m} \Pbra{i}\trho^{n+1}_s(t)\Pket{l} 
    - \delta_{n,1} e^{-i (-\e{i}{l}) s} 
    \D_{j,i} \DD_{i,l} \Pbra{l}\trho^n_s(t)\Pket{m}  \Big)  \\ \nonumber
    &+C^\nu_{00}(-s) \Big( \delta_{n,0} e^{-i \e{m}{l} s} 
    \DD_{j,i}\D_{l,m} \Pbra{i}\trho^{n+1}_{s}(t)\Pket{l}
    -  \delta_{n,1} e^{-i \e{l}{i} s} 
    \D_{i,l} \DD_{l,m} \Pbra{j}\trho^n_s(t)\Pket{i} \Big)  \\ \nonumber
    &+C^\nu_{11}(s) \Big(\delta_{n,1} e^{-i \e{i}{j} s}   
    \D_{j,i}\DD_{l,m} \Pbra{i}\trho^{n-1}_s(t)\Pket{l} 
    -  \delta_{n,0} e^{-i \e{l}{i} s} 
    \DD_{j,i}\D_{i,l} \Pbra{l} \trho^n_s(t)\Pket{m}  \Big)  \\ \label{Eq_app:Redfield_before_int_s}
    &+C^\nu_{11}(-s) \Big( \delta_{n,1} e^{-i (-\e{l}{m}) s} 
    \D_{j,i}\DD_{l,m} \Pbra{i}\trho^{n-1}_{s}(t)\Pket{l}
    -  \delta_{n,0} e^{-i (-\e{i}{l}) s} 
    \DD_{i,l}\D_{l,m} \Pbra{j}\trho^n_s(t)\Pket{i} \Big) \Bigg)
\end{align}
At this point, we are look carefully at the integration over $s$. We separate everything that contains the same integration functions/sign and use $E$ to represent all the frequency differences $\e{k}{l}$,
\begin{align}\label{Eq_app:Int_C_start}
    &\int^{+\infty}_0 \mathrm{d}s \, C^\nu_{00}(s) e^{-i E s}= \frac{1}{2\pi} \int_{-\infty}^\infty d\omega' \, \int^{+\infty}_0 \mathrm{d}s \,  \G(\omega') [1-\f(\omega' )] e^{-i (E+\omega') s} = \frac{1}{2} \G(-E) [1-\f(-E)] 
    \\  \label{Eq_app:Int_C_start-1}
    &\int^{+\infty}_0 \mathrm{d}s \,  C^\nu_{11}(s) e^{-i E s}=\frac{1}{2\pi}  \int_{-\infty}^\infty d\omega' \, \int^{+\infty}_0 \mathrm{d}s \, \G(\omega') \f(\omega') e^{-i (E-\omega') s} = \frac{1}{2} \G(E) \f(E)
    \\ \label{Eq_app:Int_C_end+1}
    &\int^{+\infty}_0 \mathrm{d}s \,  C^\nu_{00}(-s) e^{-i E s} =\frac{1}{2\pi} \int_{-\infty}^\infty d\omega' \, \int^{+\infty}_0 \mathrm{d}s \,\G(\omega') [1-\f(\omega' )] e^{-i (E-\omega') s} = \frac{1}{2} \G(E) [1-\f(E)]
    \\ \label{Eq_app:Int_C_end}
    &\int^{+\infty}_0 \mathrm{d}s \,  C^\nu_{11}(-s) e^{-i E s} =\frac{1}{2\pi}  \int_{-\infty}^\infty d\omega' \, \int^{+\infty}_0 \mathrm{d}s \,\G(\omega') \f(\omega') e^{-i (E+\omega') s} = \frac{1}{2} \G(-E) \f(-E), 
\end{align}
where in the second equality we used the explicit expression of the bath correlation functions computed in Eq.~\eqref{Eq_app:C00}-\eqref{Eq_app:C11} and in the last equality we implemented the Sokhotski–Plemelj theorem, i.e.
\begin{equation}\label{Eq_app:SPTheorem}
    \int_{-\infty}^{+\infty} d\omega' \int_{0}^{\infty} d\tau F(\omega') e^{-i(E\pm\omega')\tau}= \pi F(\mp E)-iP\int_{-\infty}^{+\infty} d\omega' \frac{F(\omega')}{E\pm\omega'}.
\end{equation}
with $P$ the principal value and $F$ an analytic function. The imaginary component is related to the Lamb-shift which we neglect~\cite{Potts2024,Strasberg2022Book}. A closer look at this component and the regime when it can be neglected is covered in Appendix~\ref{Appendix:Lamb_Shift}.
Replacing Eq.~\eqref{Eq_app:Int_C_start}-\eqref{Eq_app:Int_C_end} into Eq.~\eqref{Eq_app:Redfield_before_int_s} yields
\begin{align}\nonumber 
    \Pbra{j}\partial_t \trho^n_s(t) \Pket{m} &=  -i \omega(j-m) \Pbra{j} \trho^n_s (t) \Pket{m} \\ \nonumber
    +\frac{1}{2}\sum_{\nu} \sum_{il}\Bigg(
    &\delta_{n,0} \G(\e{j}{i}) [1-\f(\e{j}{i})]\DD_{j,i}\D_{l,m} \Pbra{i}\trho^{n+1}_s(t)\Pket{l} 
    - \delta_{n,1} \G(\e{i}{l}) [1-\f(\e{i}{l})] \D_{j,i} \DD_{i,l} \Pbra{l}\trho^n_s(t)\Pket{m}   \\ \nonumber
    &+ \delta_{n,0} \G(\e{m}{l}) [1-\f(\e{m}{l})] \DD_{j,i}\D_{l,m} \Pbra{i}\trho^{n+1}_{s}(t)\Pket{l}
    -  \delta_{n,1} \G(\e{l}{i}) [1-\f(\e{l}{i})] \D_{i,l} \DD_{l,m} \Pbra{j}\trho^n_s(t)\Pket{i} \\ \nonumber
    &+ \delta_{n,1} \G(\e{i}{j}) \f(\e{i}{j}) \D_{j,i}\DD_{l,m} \Pbra{i}\trho^{n-1}_s(t)\Pket{l} 
    -  \delta_{n,0} \G(\e{l}{i}) \f(\e{l}{i}) \DD_{j,i}\D_{i,l} \Pbra{l} \trho^n_s(t)\Pket{m} \\ 
    &+\delta_{n,1} \G(\e{l}{m}) \f(\e{l}{m}) \D_{j,i}\DD_{l,m} \Pbra{i}\trho^{n-1}_{s}(t)\Pket{l}
    -  \delta_{n,0} \G(\e{i}{l}) \f(\e{i}{l}) \DD_{i,l}\D_{l,m} \Pbra{j}\trho^n_s(t)\Pket{i}  \Bigg)
\end{align}
Next, we adjust the summation indexes in order for the elements of the density matrix to be the same.
\begin{align}\nonumber 
    \Pbra{j}\partial_t \trho^n_s(t) \Pket{m} &=  -i \omega(j-m) \Pbra{j} \trho^n_s (t) \Pket{m} \\ \nonumber
    +\frac{1}{2}\sum_{\nu} \sum_{kl} \Bigg(
    & \delta_{n,0} \G(\e{j}{k}) [1-\f(\e{j}{k})]\DD_{j,k}\D_{l,m} \Pbra{k}\trho^{n+1}_s(t)\Pket{l} 
    - \sum_{i}\delta_{n,1} \G(\e{i}{k}) [1-\f(\e{i}{k})] \D_{j,i} \DD_{i,k} \Pbra{k}\trho^n_s(t)\Pket{m} \delta_{ml}  \\ \nonumber
    &+  \delta_{n,0} \G(\e{m}{l}) [1-\f(\e{m}{l})] \DD_{j,k}\D_{l,m} \Pbra{k}\trho^{n+1}_{s}(t)\Pket{l}
    - \sum_{i} \delta_{n,1} \G(\e{i}{l}) [1-\f(\e{i}{l})] \D_{l,i} \DD_{i,m} \delta_{jk}\Pbra{k}\trho^n_s(t)\Pket{l} \\ \nonumber
    &+ \delta_{n,1} \G(\e{k}{j}) \f(\e{k}{j}) \D_{j,k}\DD_{l,m} \Pbra{k}\trho^{n-1}_s(t)\Pket{l} 
    - \sum_{i} \delta_{n,0} \G(\e{k}{i}) \f(\e{k}{i}) \DD_{j,i}\D_{i,k} \Pbra{k} \trho^n_s(t)\Pket{m} \delta_{ml}  \\ \label{Eq_app:Redfield_before_tensors}
    &+\delta_{n,1} \G(\e{l}{m}) \f(\e{l}{m}) \D_{j,k}\DD_{l,m} \Pbra{k}\trho^{n-1}_{s}(t)\Pket{l} 
    - \sum_{i} \delta_{n,0} \G(\e{l}{i}) \f(\e{l}{i}) \DD_{l,i}\D_{i,m} \delta_{kj}\Pbra{k}\trho^n_s(t)\Pket{l} \Bigg)
\end{align}
From here, we define the \textit{Redfield tensors} corresponding to lead $\nu$ based on the different transition of the electronic state as
\begin{align}\nonumber
    \Rnn{n,}{n}{j}{m}{k}{l} & =  \frac{1}{2} \sum_i \Bigg[ 
    \delta_{n,0}
    \Big( \G(\e{k}{i}) \f(\e{k}{i}) \DD_{j,i}\D_{i,k} \delta_{ml} +  \G(\e{l}{i}) \f(\e{l}{i}) \DD_{l,i}\D_{i,m} \delta_{kj} \Big) \\ \label{Eq_app:A_nn_}
    &\quad \quad \quad + \delta_{n,1}\Big( \G(\e{i}{k}) [1-\f(\e{i}{k})] \D_{j,i} \DD_{i,k}\delta_{ml} + \G(\e{i}{l}) [1-\f(\e{i}{l})] \D_{l,i} \DD_{i,m} \delta_{jk} \Big) \Bigg]\\ \label{Eq_app:A_nn+1}
    \Rnn{n+1,}{n}{j}{m}{k}{l} &= \frac{1}{2} \delta_{n,0} \DD_{j,k}\D_{l,m} \Big(\G(\e{j}{k}) [1-\f(\e{j}{k})]+ \G(\e{m}{l}) [1-\f(\e{m}{l})]  \Big)
    \\ \label{Eq_app:A_nn_1}
    \Rnn{n-1,}{n}{j}{m}{k}{l} &= \frac{1}{2} \delta_{n,1} \D_{j,k}\DD_{l,m}  \Big( \G(\e{k}{j}) \f(\e{k}{j}) + \G(\e{l}{m}) \f(\e{l}{m}) \Big)
\end{align}
Replacing these definitions into Eq.~\eqref{Eq_app:Redfield_before_tensors} we arrive to the projected second order Redfield master equation
\begin{equation}\label{Eq_app:ME_Redfield_final}
    \Pbra{j}\partial_t\trho_s^n \Pket{m} = -i\omega(j-m)\Pbra{j} \trho_s^n \Pket{m} + \sum_\nu \sum_{kl}
    \Big( -\Rnn{n,}{n}{j}{m}{k}{l} \Pbra{k} \trho_s^n \Pket{l}
    + \Rnn{n+1,}{n}{j}{m}{k}{l} \Pbra{k} \trho_s^{n+1} \Pket{l}
     + \Rnn{n-1,}{n}{j}{m}{k}{l} \Pbra{k} \trho_s^{n-1} \Pket{l} \Big).    
\end{equation}
which corresponds to Eq.~\eqref{Eq:ME_Redfield_final} of the main text. The non-equilibrium steady state is obtained by taking $\Pbra{j}\trho_s^n \Pket{m} =0 \, \forall \, j,m,n$.

\subsection{GKLS form}\label{Appendix:LindblatForm}

In this section we make an additional approximation (the secular approximation) over Eq.~\eqref{Eq_app:Redfield_simple_Interaction}  in order to bring it to GKLS form. Before this, we write the system operators in the interaction picture in a Fourier series~\cite{Potts2024} 
\begin{align}\label{Eq_app:system_fourier}
    \Int{(\S_{0})} (t)  = \sum_\alpha e^{-i\omega_\alpha t} \S^\alpha_{0} 
    \qquad\qquad \Int{(\S_{1})} (t) = \sum_j e^{-i\omega_\alpha t} \tilde{S}^\alpha_{1}, 
\end{align}
where we have made use of the the eigenbasis of the system Hamiltonian under the polaron transformation (see Eq.~\eqref{eq:Energy_system}) such that 
\begin{align}\label{Eq_app:system_int_S0}
    \S^{\{{i,\tau},{l,\sigma}\}}_{0} &= \Fket{i}{\tau} \Fbra{i}{\tau} \S_{0} \Fket{l}{\sigma}\Fbra{l}{\sigma}=
    \sqrt{\sigma}  \delta_{\tau,\sigma-1} \DD_{il} \Fket{i}{\tau}\Fbra{l}{\sigma} \\ \label{Eq_app:system_int_S1}
    \S^{\{ {i,\tau},{l,\sigma}\}}_{1} &= \Fket{i}{\tau} \Fbra{i}{\tau} \S_{1} \Fket{l}{\sigma} \Fbra{l}{\sigma} = \sqrt{\tau} \delta_{\tau-1,\sigma} \D_{il} \Fket{i}{\tau}\Fbra{l}{\sigma}, 
\end{align}
in the following we will make use of the short-hand notation $\alpha=\{{i,\tau},{l,\sigma}\}$ and 
\begin{equation}\label{Eq_app:omega_j_notation}
    \omega_\alpha= \tilde{E}_{l,\sigma}-\tilde{E}_{i,\tau} = \omega (l-i) +\tilde{\mu} (\sigma-\tau). 
\end{equation}
We remind the reader that $\tilde{E}_{j,\tau}$ is the energy of the system under the polaron frame as defined in Eq.~\eqref{eq:Energy_system} of the main text. Thus replacing Eq.~\eqref{Eq_app:system_fourier} into Eq.~\eqref{Eq_app:Redfield_simple_Interaction} and using that the only no-zero correlation functions are Eq.~\eqref{Eq_app:C00} and Eq.~\eqref{Eq_app:C11} we obtain~\cite{Potts2024,Strasberg2022Book}
\begin{equation}\label{Eq_app:GKLS_before_secular}
    \partial_t\Int{(\trho_{s})}(t) =\sum_{\nu} \sum_{q=0,1}\sum_{\alpha,\alpha'} e^{i(\omega_\alpha-\omega_{\alpha'})t} \,  \Omega^\nu_{q}(\omega _{\alpha'})
    \Big( \S^{\alpha'}_{q} \,  \Int{(\trho_{s})}(t) \, \big(\S^\alpha_{q} \big)^\dagger- 
    \big(\S^\alpha_{q}\big)^\dagger \, \S^{\alpha'}_{q} \, \Int{(\trho_{s})}(t) \Big) +h.c.
\end{equation}
with
\begin{equation}\label{Eq_app:G_defintion}
    \Omega^\nu_{q}(\omega)= \Re\left( \int_0^{+\infty} ds e^{i\omega s} C^\nu_{qq}(s) \right),  \quad q=0,1.
\end{equation}
Which we have solved explicitly in Eq.~\eqref{Eq_app:Int_C_start} for $q=0$ and Eq.~\eqref{Eq_app:Int_C_start-1} for $q=1$. We note that we only take the real part of the solution as the imaginary component is related to the Lamb-shift which we have neglected, see Appendix~\ref{Appendix:Lamb_Shift} for further details. 

At this point, we take the secular approximation: $e^{i(\omega_\alpha-\omega_{\alpha'})t}$ oscillates quickly compared to the time evolution induced by $\Omega^\nu_{k}$ which is valid for~\cite{Strasberg2022Book,Potts2024}
\begin{equation}\label{Eq_app:secular_aprox}
    |\Omega^\nu_{q}(\omega_\alpha)| \ll |\omega_\alpha-\omega_{\alpha'}|  \quad \forall \, \omega_\alpha\neq \omega_{\alpha'} .
\end{equation}
If this condition is met one can take $e^{i(\omega_\alpha-\omega_{\alpha'})t}\sim \delta_{\alpha,\alpha'} $. In Appendix~\ref{Appendix:Secular_aprox_cond} we study how this condition results in a restriction over the physical parameters. There we conclude that for Eq.~\eqref{Eq_app:secular_aprox} to be satisfied, the parameters have to satisfy the following "loose bound"
\begin{equation}
     \Gamma_\nu \ll 2 \max\{\tilde{\mu},\omega\}, \quad  \, \nu=L,R.
\end{equation}
Thus applying this approximation to Eq.~\eqref{Eq_app:GKLS_before_secular} and returning to the Schrödinger picture yields~\cite{Potts2024}
\begin{equation}\label{Eq_app:GKLS_after_secular}
    \partial_t \trho_{s}(t) = -i [\tHs,\trho_s(t)]
    + \sum_{\nu} \sum_{\tau,\sigma} \sum_{i,l} \Omega^\nu_0(\omega_{\{\tau,i,\sigma,l\}}) \mathcal{D}[\S^{\{\tau,i,\sigma,l\}}_{ 0}] \trho_s(t) + \Omega^\nu_1(\omega_{\{\tau,i,\sigma,l\}}) \mathcal{D}[\S^{\{\tau,i,\sigma,l\}}_{1}] \trho_s(t).
\end{equation}
where $\mathcal{D}[\hat{A}]\rho= \hat{A}\rho \hat{A}^\dagger -\frac{1}{2}\{\hat{A}^\dagger \hat{A}, \rho \} $ are the dissipation super-operators.
Similar to the Redfield equation detailed in Appendix~\ref{Appendix:Redfield}, we project the density matrix into the energy eigenbasis, see Eq.~\eqref{eq:Energy_system} of the main text, i.e.
\begin{align}\nonumber 
    \Pbra{j}\partial_t \trho^n_s(t) \Pket{m} & =  -i \, \omega( j - m) \Pbra{j} \trho^n_s (t) \Pket{m} 
    \\ \label{Eq_app:GKLS_after_secular_projection}
    + & \sum_{\nu} \sum_{\tau,\sigma} \sum_{i,l} 
    \Omega^\nu_0(\omega_{\{i,\tau,l,\sigma\}})
    \underbrace{\Fbra{j}{n} \mathcal{D}[\S^{\{i,\tau,l,\sigma\}}_{ 0}] \trho_s(t) \Fket{m}{n}}_{\text{(\textbf{I})}} +
    \Omega^\nu_1(\omega_{\{i,\tau,l,\sigma\}})
    \underbrace{\Fbra{j}{n} \mathcal{D}[\S^{\{i,\tau,l,\sigma\}}_{1}] \trho_s(t) \Fket{j}{n}}_{\text{(\textbf{II})}}.
\end{align}
In the following we compute the quantities labeled as \textbf{I} and \textbf{II} separately; 
\begin{align} \nonumber
    \text{(\textbf{I})}&= \Fbra{j}{n}  \mathcal{D}[\S^{\{i,\tau,l,\sigma\}}_{ 0}] \trho_s \Fket{m}{n} 
    = \Fbra{j}{n} \S^{\{i,\tau,l,\sigma\}}_{0} \trho_s \big( \S^{\{i,\tau,l,\sigma\}}_{0}\big) ^\dagger \Fket{m}{n} -\frac{1}{2} \Fbra{j}{n} \{\big(\S^{\{i,\tau,l,\sigma\}}_{0}\big)^\dagger \S^{\{i,\tau,l,\sigma\}}_{0}, \trho_s \} \Fket{m}{n} \\ \nonumber
    &= \Fbra{j}{n} \S^{\{i,\tau,l,\sigma\}}_{0} \trho_s \S^{\{l,\sim,i,\tau\}}_{1} \Fket{m}{n} -\frac{1}{2} \Fbra{j}{n} \{\S^{\{l,\sigma,i,\tau\}}_{1} \S^{\{i,\tau,l,\sigma\}}_{0}, \trho_s \} \Fket{m}{n} \\ \nonumber
    &= \Fbra{j}{n} \S^{\{i,\tau,l,\sigma\}}_{0} \trho_s \S^{\{l,\sigma,i,\tau\}}_{1} \Fket{m}{n} - \frac{1}{2} \Big( \Fbra{j}{n} \S^{\{l,\sigma,i,\tau\}}_{1} \S^{\{i,\tau,l,\sigma\}}_{0} \trho_s \Fket{m}{n} + \Fbra{j}{n} \trho_s \S^{\{l,\sigma,i,\tau\}}_{1} \S^{\{i,\tau,l,\sigma\}}_{0} \Fket{m}{n} \Big) \\\nonumber
    &= \sigma \delta_{\tau,\sigma-1} \DD_{il} \D_{li} \Big(   \delta_{j,i} \delta_{i,m} \delta_{n,\tau}\Fbra{l}{\sigma} \trho_s \Fket{l}{\sigma} - \frac{1}{2} \Big( \delta_{j,l}\delta_{n,\sigma} \Fbra{l}{\sigma} \trho_s \Fket{m}{n} + \Fbra{j}{n} \trho_s \Fket{l}{\sigma} \delta_{l,m}\delta_{\sigma,n} \Big) \Big) \\ \label{Eq_app:Quantity_(I)}
    &= \delta_{\sigma,1} \delta_{\tau,0} \DD_{il} \D_{li} \Big(\delta_{n,0} \delta_{j,i} \delta_{i,m} \Pbra{l} \trho^1_s \Pket{l} - \frac{1}{2} \delta_{1,n}  \Big( \delta_{j,l} \Pbra{l}\trho^1_s \Pket{m} + \Pbra{j}\trho^1_s \Pket{l} \delta_{l,m}\Big) \Big) 
\end{align}
\begin{align}\nonumber
    \text{(\textbf{II})}&= \Fbra{j}{n} \mathcal{D}[\S^{\{i,\tau,l,\sigma\}}_{1}] \trho_s \Fket{m}{n} 
    = \Fbra{j}{n} \S^{\{i,\tau,l,\sigma\}}_{1} \trho_s \big(\S^{\{i,\tau,l,\sigma\}}_{1}\big)^\dagger\Fket{m}{n}  -\frac{1}{2} \Fbra{j}{n} \{\big(\S^{\{ i,\tau,l,\sigma \}}_{1}\big)^\dagger \S^{\{i,\tau,l,\sigma\}}_{1}, \trho_s \} \Fket{m}{n} \\ \nonumber
    &= \Fbra{j}{n} \S^{\{i,\tau,l,\sigma\}}_{1} \trho_s \S^{\{l,\sigma,i,\tau\}}_{0} \Fket{m}{n} -\frac{1}{2} \Fbra{j}{n} \{\S^{\{l,\sigma,i,\tau\}}_{0} \S^{\{i,\tau,l,\sigma\}}_{1}, \trho_s \} \Fket{m}{n} \\ \nonumber
    &= \Fbra{j}{n} \S^{\{i,\tau,l,\sigma\}}_{1} \trho_s \S^{\{l,\sigma,i,\tau\}}_{0} \Fket{m}{n} 
    -\frac{1}{2} \Big( \Fbra{j}{n} \S^{\{l,\sigma,i,\tau\}}_{0} \S^{\{i,\tau,l,\sigma\}}_{1} \trho_s  \Fket{m}{n} + \Fbra{j}{n} \trho_s  \S^{\{l,\sigma,i,\tau\}}_{0} \S^{\{i,\tau,l,\sigma\}}_{1} \Fket{m}{n} \Big) \\ \nonumber
    &= \tau \delta_{\tau-1,\sigma} \D_{il} \DD_{li} \Big( \delta_{j,i}\delta_{i,m}\delta_{n,\tau} \Fbra{l}{\sigma} \trho_s \Fket{l}{\sigma} 
    -\frac{1}{2} \Big( \delta_{j,l}\delta_{n,\sigma} \Fbra{l}{\sigma} \trho_s  \Fket{m}{n} 
    + \delta_{n,\sigma} \delta_{l,m} \Fbra{j}{n} \trho_s  \Fket{l}{\sigma} \Big) \Big) \\ \label{Eq_app:Quantity_(II)}
    &= \delta_{\tau,1} \delta_{0,\sigma} \D_{il} \DD_{li} \Big( \delta_{n,1}  \delta_{j,i}\delta_{i,m}\Pbra{l}\trho^0_s \Pket{l} 
    -\frac{1}{2} \delta_{n,0}\Big( \delta_{j,l} \Pbra{l} \trho^0_s \Pket{m} + \delta_{l,m} \Pbra{j} \trho^0_s  \Pket{l}\Big) \Big) .
\end{align}
Replacing Eq.~\eqref{Eq_app:Quantity_(I)}-\eqref{Eq_app:Quantity_(II)} into Eq.~\eqref{Eq_app:GKLS_after_secular_projection} yields
\begin{align}\nonumber 
    \Pbra{j}\partial_t \trho^n_s \Pket{m} & =  -i \, \omega( j - m) \Pbra{j} \trho^n_s  \Pket{m} 
    \\ \nonumber
    +  \sum_{\nu} \sum_{i,l} 
    & \Omega^\nu_0(\omega_{\{i,0,l,1\}})
    \DD_{il} \D_{li} \Big(\delta_{n,0} \delta_{j,i} \delta_{i,m} \Pbra{l} \trho^{n+1}_s \Pket{l} - \frac{1}{2} \delta_{n,1}  \Big( \delta_{j,l} \Pbra{l}\trho^n_s \Pket{m} + \Pbra{j}\trho^n_s \Pket{l} \delta_{l,m}\Big) \Big) \\
    &+
    \Omega^\nu_1(\omega_{\{i,1,l,0\}})
    \D_{il} \DD_{li} \Big( \delta_{n,1}  \delta_{j,i}\delta_{i,m}\Pbra{l}\trho^{n-1}_s \Pket{l} 
    -\frac{1}{2} \delta_{n,0}\Big( \delta_{j,l} \Pbra{l} \trho^n_s \Pket{m} + \delta_{l,m} \Pbra{j} \trho^n_s  \Pket{l}\Big) \Big).
\end{align}
Finally, by using the shorthand notation $\omega_{i,j}$ and the explicit expressions of $\Omega_0^\nu/\Omega_1^\nu$ and arranging some indexes we arrive to the final projected master equation of GKLS form
\begin{align}\nonumber 
    \Pbra{j}\partial_t \trho^n_s \Pket{m}  =  -i \, \omega( j - m) \Pbra{j} \trho^n_s & \Pket{m} +\frac{1}{2} \sum_{\nu} \sum_{l} \Big( \Rnn{n+1,}{n}{j}{j}{l}{l}
    \delta_{j,m} \Pbra{l} \trho^{n+1}_s \Pket{l} + 
    \Rnn{n-1,}{n}{j}{j}{l}{l}
    \delta_{j,m} \Pbra{l}\trho^{n-1}_s \Pket{l} \\ \label{Eq_app:Lindblat_projected}
   & -  \frac{1}{2}   \Big( \Rnn{n-1,}{n}{l}{l}{j}{j} + \Rnn{n-1,}{n}{l}{l}{m}{m} \Big) \Pbra{j}\trho^n_s \Pket{m}  - \frac{1}{2}   
    \Big( \Rnn{n+1,}{n}{l}{l}{j}{j}   + \Rnn{n+1,}{n}{l}{l}{m}{m} \Big) \Pbra{j} \trho^n_s \Pket{m} \Big).
\end{align}

\section{In depth analysis of the Lamb-shift}\label{Appendix:Lamb_Shift}

In this section we explicitly compute Eq.~\eqref{Eq_app:Int_C_start}-\eqref{Eq_app:Int_C_end} and study the limitations on the parameters in order for Eq.~\eqref{Eq:ME_Redfield_final} of the main text to be valid. 

\begin{figure}[ht]
    \centering
    \includegraphics[width=.7\linewidth]{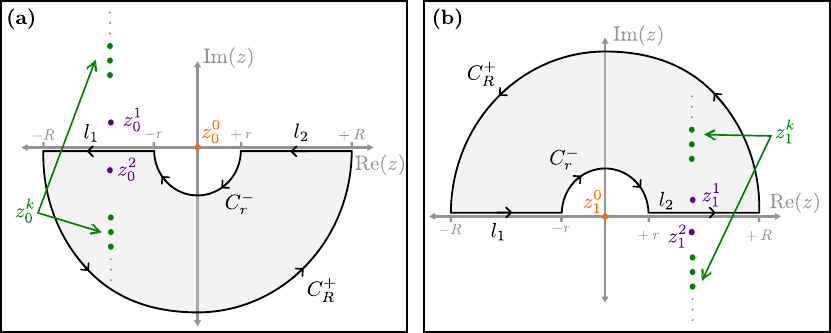}
    \caption{\textbf{Integration curves used to compute the Lamb-shift.}
    Panel \textbf{(a)} corresponds to the curve used for Eq.~\eqref{Eq_app:residue_eq_g0} while panel \textbf{(b)} was used for Eq.~\eqref{Eq_app:residue_eq_g1}. For simplicity, in the graphics we consider $E\geq0$.}
    \label{fig:Figure1_appendix}
\end{figure}

We re-write Eq.~\eqref{Eq_app:Int_C_start}-\eqref{Eq_app:Int_C_end} using the following two functions 
\begin{align}\label{Eq_app:def_G0}
    G^\nu_0(E) &= \frac{1}{2\pi} \int_{-\infty}^{+\infty} \mathrm{d}\omega' \int_0^{+\infty} \mathrm{d}s \, g_0(\omega') e^{-i(-E \pm\omega')s} = \frac{1}{2}  g_0(\mp E)  \mp i P.V. \int_{-\infty}^{+\infty} \frac{\mathrm{d}\omega'}{2\pi} \frac{g_0(\omega')}{\omega'\pm E}  \\ \label{Eq_app:def_G1}
    G^\nu_1(E) &= \frac{1}{2\pi} \int_{-\infty}^{+\infty} \mathrm{d}\omega' \int_0^{+\infty} \mathrm{d}s \, g_1(\omega') e^{-i(-E \pm\omega')s} =\frac{1}{2}  g_1(\mp E) \mp i P.V \int_{-\infty}^{+\infty} \frac{\mathrm{d}\omega'}{2\pi} \frac{g_1(\omega')}{\omega'\pm E}
\end{align}
and retain the imaginary part that arises from the Sokhotski–Plemelj theorem. We have taken for simplicity $g_0(E)=\G(E) [1-\f(E)]$ and $g_1(E)=\G(E) \f(E)$. Note that Eq.~\eqref{Eq_app:def_G0} corresponds to Eq.~\eqref{Eq_app:Int_C_start} with the plus sign and Eq.~\eqref{Eq_app:Int_C_end+1} with the minus sign, while Eq.~\eqref{Eq_app:def_G1} corresponds to the Eq.~\eqref{Eq_app:Int_C_end} with plus sign and Eq.~\eqref{Eq_app:Int_C_start-1} with the minus sign. The difficulty arises in computing
\begin{align}\label{Eq_app:def_G0_Im}
    \Im(G^\nu_0(E)) &= \mp P.V. \int_{-\infty}^{+\infty} \frac{\mathrm{d} x}{2\pi} \frac{g_0(x\mp E)}{x}  \\ \label{Eq_app:def_G1_Im}
    \Im(G^\nu_1(E)) &= \mp  P.V \int_{-\infty}^{+\infty} \frac{\mathrm{d}x}{2\pi} \frac{g_1(x\mp E)}{x}
\end{align}
where we have taken the change of variables $x= \omega' \pm E $.
Here we carry out this computation using the Residue theorem. For this, the first step consists on taking the integration evaluated in the complex plane as
\begin{align}
    \frac{g_0(x\mp E)}{x}&\rightarrow \frac{g_0(z\mp E)}{z}= \frac{1}{z} \frac{\Gamma_\nu \delta_\nu^2}{(z\mp E -\gamma_\nu)+\delta_\nu^2} \frac{1}{e^{-\beta_\nu(z\mp E -\mu_\nu)}+1} \\ 
    \frac{g_1(x\mp E)}{x}&\rightarrow \frac{g_1(z\mp E)}{z} = \frac{1}{z} \frac{\Gamma_\nu \delta_\nu^2}{(z\mp E -\gamma_\nu)+\delta_\nu^2} \frac{1}{e^{\beta_\nu(z\mp E -\mu_\nu)}+1} .
\end{align} 
where we have written the explicit form of the Lorentzian spectral density defined in Eq.~\eqref{eq:spectral_Lorentz} of the main text and the Fermi functions.
From here it is straightforward to note that the singularities of each function are 
\begin{align}\label{Eq_app:Sing_g0}
    \text{sing.}\left(\frac{g_0}{z} \right)  &=\{z^0_0=0,
    z^1_0= i\delta_\nu \pm E +\gamma_\nu, 
    z^2_0= -i\delta_\nu \pm E +\gamma_\nu, 
    z^k_0=-i\pi (2k+1)/\beta_\nu \pm E +\mu_\nu, \, k \in \mathrm{Z}\}\\ \label{Eq_app:Sing_g1}
    \text{sing.}\left(\frac{g_1}{z}\right) &=\{z^0_1=0, 
    z^1_1= i\delta_\nu \pm E +\gamma_\nu , 
    z^2_1=- i\delta_\nu \pm E+\gamma_\nu , 
    z^k_1= i\pi (2k+1)/\beta_\nu \pm E+\mu_\nu, \,  k \in \mathrm{Z} \}
\end{align}
In Fig.~\ref{fig:Figure1_appendix} we plot each one along the complex plane. The first one labeled as $z^0_{0/1}$ arises form the factor $1/z$. The second and third arise labels as $z^1_{0/1}$ and $z^2_{0/1}$ respectively arise from the Lorentz shape of the spectral density $\G$ defined in Eq.~\eqref{eq:spectral_Lorentz} of the main text. Here we note that if the typical \textit{wide-band limit} was chosen, there will not be any singularity arising from the spectral densities. Also, in the main text it was mentioned that this work can easily be easily generalized to an arbitrary spectral density by taking a sum of Lorentz spectral densities, which is true as long as any new singularities is properly considered. Finally the remaining singularities, $z^k_{0/1}$, arise from the Fermi functions.  
Now, we integrate $g_0/z$ over the curve $C_0$ oriented positive as shown in Fig.~\subfigref{fig:Figure1_appendix}{a} and $C_1$ as shown in Fig.~\subfigref{fig:Figure1_appendix}{b} for $g_1/z$. We can separate our integration by smaller segments 
\begin{align} \label{Eq_app:residue_eq_g0}
    \int_{C_0^+} \mathrm{d}z \, \frac{g_0(z\mp E)}{z} &= \int_{l_1+l_2} \mathrm{d}z \, \frac{g_0(z\mp E)}{z}  - \int_{C_r^+}  \mathrm{d}z \, \frac{g_0(z\mp E)}{z} + \int_{C_R^+} \mathrm{d}z \, \frac{g_0(z\mp E)}{z}= 2\pi i \sum_n \text{Res}\left(\frac{g_0}{z},z^n_0\right) . \\
    \label{Eq_app:residue_eq_g1}
    \int_{C_1^+} \mathrm{d}z \, \frac{g_1(z\mp E)}{z}  &= \int_{l_1+l_2}  \mathrm{d}z \, \frac{g_1(z\mp E)}{z} - \int_{C_r^+} \mathrm{d}z \,  \frac{g_1(z\mp E)}{z} + \int_{C_R^+} \mathrm{d}z \, \frac{g_1(z\mp E)}{z}= 2\pi i \sum_n \text{Res}\left(\frac{g_1}{z},z^n_1 \right) . 
\end{align}
as indicated in Fig.~\ref{fig:Figure1_appendix}\textbf{(a)} for Eq.~\eqref{Eq_app:residue_eq_g0} and Fig.~\ref{fig:Figure1_appendix}\textbf{(b)} for Eq.~\eqref{Eq_app:residue_eq_g1} respectively. Where in the last equality we have applied the Residue theorem where $z^n_0/z^n_1$ are all the singularities inside the curve $C_0$ or $C_1$ depending on the case.

At this point we remind the reader that the task in this section is to compute $\Im(G^\nu_0(E))$ (Eq.~\eqref{Eq_app:def_G0_Im}) and $\Im(G^\nu_1(E))$ (Eq.~\eqref{Eq_app:def_G1_Im}). We recover these expressions with the integration over the real axis as,
\begin{align}\label{Eq_app:real_integration_g0}
    \lim_{\substack{{r\rightarrow 0}\\{R\rightarrow -\infty}}} \int_{l_1+l_2} \mathrm{d}z\, \frac{g_0(z\mp E)}{z} &= P.V. \int_{-\infty}^{+\infty} \mathrm{d}x\, \frac{g_0(x\mp E)}{x} = 2\pi \Im(G^\nu_0(E)) \\ \label{Eq_app:real_integration_g1}
    \lim_{\substack{{r\rightarrow 0}\\{R\rightarrow +\infty}}} \int_{l_1+l_2} \mathrm{d}z\, \frac{g_1(z\mp E)}{z} &= P.V. \int_{-\infty}^{+\infty} \mathrm{d}x\, \frac{g_1(x\mp E)}{x} = 2\pi \Im(G^\nu_1(E)).
\end{align}
Thus combining Eq.~\eqref{Eq_app:residue_eq_g0}-\eqref{Eq_app:real_integration_g1} we obtain 
\begin{align}\label{Eq_app:G0_residual}
    \Im(G^\nu_0(E)) &= \frac{1}{2\pi} \lim_{\substack{{r\rightarrow 0}\\{R\rightarrow +\infty}}} \left( 2\pi i \sum_n \text{Res} \left(\frac{g_1}{z},z_n \right) + \int_{C_r^+} \mathrm{d}z\, \frac{g_0(z\mp E)}{z} - \int_{C_R^+}  \mathrm{d}z\, \frac{g_0(z\mp E)}{z} \right) \\\label{Eq_app:G1_residual}
    \Im(G^\nu_1(E)) &= \frac{1}{2\pi}\lim_{\substack{{r\rightarrow 0}\\{R\rightarrow -\infty}}}  \left( 2\pi i \sum_n \text{Res}\left(\frac{g_1}{z},z_n \right) + \int_{C_r^+}  \mathrm{d}z\, \frac{g_1(z\mp E)}{z} - \int_{C_R^+}  \mathrm{d}z\, \frac{g_1(z\mp E)}{z} \right).
\end{align}
Such that our original task is dramatically simplified to computing residuals and path integrals where theorems can be readily used. We compute each component separately as follows;

We start with the computation of the residues. By observing Fig.~\ref{fig:Figure1_appendix}, it is straightforward to note that for each case the singularities inside the cure are
\begin{align}\label{Eq_app:Residual_comp_start}
    \sum_n \text{Res}\left(\frac{g_0}{z},z^n_0\right) &= \text{Res}\left(\frac{g_0}{z},z^2_0\right) + \sum_{k\geq 0} \text{Res}\left(\frac{g_0}{z},z^k_0\right)\\
    \sum_n \text{Res}\left(\frac{g_1}{z},z^n_0\right) &= \text{Res}\left(\frac{g_1}{z},z^1_1\right) + \sum_{k\geq 0} \text{Res}\left(\frac{g_1}{z},z^k_1\right).
\end{align}
Using the fact that $z_0^2,z_0^k,z_1^1$ and $z_1^k$ are simple poles, the computation of each residue is 
\begin{align}
    \text{Res}\left(\frac{g_0}{z},z^2_0\right) &= \lim_{z\rightarrow z_0^2} (z-z_0^2) \frac{g_0(z\mp E)}{z} =  \frac{i\Gamma_\nu \delta_\nu }{2 z_0^2} [1-\f (z_0^2 \mp E)] \\
    \text{Res}\left(\frac{g_1}{z},z^1_1\right) &= \lim_{z\rightarrow z_1^1} (z-z_1^1) \frac{g_1(z\mp E)}{z} =  \frac{-i\Gamma_\nu \delta_\nu }{2 z_1^1} \f (z_1^1 \mp E) \\
    \text{Res}\left(\frac{g_0}{z},z^k_0\right) &= \lim_{z\rightarrow z^k_0} (z-z^k_0) \frac{g_0(\mp E)}{z} = \frac{+1}{\beta_\nu} \frac{\G (z^k_0\mp E)}{z^k_0} \\
    \text{Res}\left(\frac{g_1}{z},z^k_1\right) &= \lim_{z\rightarrow z^k_1} (z-z^k_1) \frac{g_1(z\mp E)}{z} = \frac{-1}{\beta_\nu} \frac{\G (z^k_1\mp E)}{z^k_1}. 
\end{align}
Where in the last two lines we implemented L'Hôpital's rule. Next, we continue computing the integrals over the curve $C^+_r$;
\begin{align}
    \int_{C_r^+} \mathrm{d}z\, \frac{g_0(z\mp E)}{z} &= i \pi \text{Res}\left(\frac{g_0}{z},0 \right) = i \pi \G (\mp E) [1-\f (\mp E)] \\
    \int_{C_r^+} \mathrm{d}z\, \frac{g_1(z\mp E)}{z} &= i \pi \text{Res}\left(\frac{g_1}{z},0 \right) = i \pi \G (\mp E) \f (\mp E)
\end{align}
Finally we finish computing Eq.~\eqref{Eq_app:G0_residual} and Eq.~\eqref{Eq_app:G1_residual} with integrals over the curve $C^+_R$, here we already take the limit over the radius $R$
\begin{align}
    \lim_{R\rightarrow -\infty} \int_{C_R^+}  \mathrm{d}z\, \frac{g_0(z\mp E)}{z} &= 
    \lim_{R\rightarrow -\infty} \int_{0}^\pi \mathrm{d}t\,  \frac{g_0(Re^{it}\mp E)}{Re^{it}} \, iRe^{it} =  i \int_{0}^\pi  \mathrm{d}t\,
    \lim_{R\rightarrow -\infty} \Big(\G (Re^{it}\mp E) [1-\f (Re^{it}\mp E)] \Big) =0\\ \label{Eq_app:Residual_comp_end}
    \lim_{R\rightarrow +\infty} \int_{C_R^+}  \mathrm{d}z\, \frac{g_1(z\mp E)}{z} &= \lim_{R\rightarrow +\infty} \int_{0}^\pi \mathrm{d}t \, \frac{g_1(Re^{it}\mp E)}{Re^{it}} iRe^{it} =  i \int_{0}^\pi  \mathrm{d}t\,
    \lim_{R\rightarrow -\infty} \Big(\G (Re^{it}\mp E) \f (Re^{it}\mp E) \Big) =0
\end{align}
With this we conclude the computaiton of each term of Eq.~\eqref{Eq_app:G0_residual} and Eq.~\eqref{Eq_app:G1_residual}. Now what is left is to Replace Eq.~\eqref{Eq_app:Residual_comp_start}-\eqref{Eq_app:Residual_comp_end} into Eq.~\eqref{Eq_app:G0_residual} and Eq.~\eqref{Eq_app:G1_residual} as needed which results in 
\begin{align}\label{Eq_app:G0_residual_final}
    \Im(G^\nu_0(E))  &= 
    \frac{1}{2\pi} \left(-\pi \Gamma_\nu \delta_\nu  \frac{[1-\f (-i\delta_\nu+\mu_\nu)]}{z_0^2} + \frac{2\pi i}{\beta_\nu}   \sum_{k\ge0} \frac{1}{z_0^k} \G \Big(\frac{-i\pi(2k+1)}{\beta_\nu}+\mu_\nu \Big)  + i \pi \G (\mp E) [1-\f (\mp E)] \right)
    \\\label{Eq_app:G1_residual_final}
    \Im(G^\nu_1(E))  &= 
    \frac{1}{2\pi} \left( \pi \Gamma_\nu \delta_\nu  \frac{\f (i\delta_\nu+\mu_\nu)}{z^1_1} -\frac{2\pi i}{\beta_\nu}   \sum_{k\ge0}  \frac{1}{z_1^k} \G \Big(\frac{i\pi(2k+1)}{\beta_\nu}+\mu_\nu \Big)  +  \pi i  \G (\mp E) \f (\mp E)  \right)
\end{align}
where we have used the explicit expressions of $z_0^2,z_1^1,z_0^k$ and $z_1^k$ in the numerators.

At this point we remind the reader that we explicitly consider Lorentzian spectral densities, see Eq.~\eqref{eq:spectral_Lorentz} of the main text. Thus the sum in the second term of Eq.~\eqref{Eq_app:G0_residual_final} and~\eqref{Eq_app:G1_residual_final} written explicitly is 
\begin{align}\label{Eq_app:G0_Lorentz}
    \sum_{k\ge0} \frac{1}{z_0^k} \G \Big(\frac{-i\pi(2k+1)}{\beta_\nu}+\mu_\nu \Big)  &= \sum_{k\ge0} \frac{1}{z_0^k} \frac{\Gamma_\nu\delta_\nu}{-(\pi(2k+1)/\beta_\nu)^2+\delta_\nu^2}  \\\label{Eq_app:G1_Lorentz}
    \sum_{k\ge0} \frac{1}{z_1^k} \G \Big(\frac{i\pi(2k+1)}{\beta_\nu}+\mu_\nu \Big) &= \sum_{k\ge0} \frac{1}{z_1^k} \frac{\Gamma_\nu\delta_\nu}{-(\pi(2k+1)/\beta_\nu)^2+\delta_\nu^2}.
\end{align}
Not that if $k$ is such that $\delta_\nu=\pi(2k+1)/ \beta_\nu)$ the both Eq.~\eqref{Eq_app:G0_Lorentz} and Eq.~\eqref{Eq_app:G1_Lorentz} diverges yielding a problem for our master equation approach. Thus we ask for $\delta_\nu \neq -(\pi(2k+1/\beta_\nu) \, \forall k\geq 0$ that is always satisfied. This is always fulfilled if we take 
\begin{equation}\label{Eq_app:Temp_delta_condition}
   \beta_\nu \delta_\nu <\pi .
\end{equation}
This condition can be more fined tuned, but it is outside of the scope of this work. In a future direction it would be interesting to study the physical consequences fo these conditions over the parameters. We note that if would have taken the \textit{wide-band limit} from the start, said divergence would have imposed conditions over the parameters.

\subsection{Bound on the Lamb shift}

In this section we study what are the conditions over the physical parameters such that the Lamb shift computed in Eq.~\eqref{Eq_app:G0_residual_final} and Eq.~\eqref{Eq_app:G1_residual_final} can be neglected. In other words we want to know when $|\Im(G^\nu_{0/1}(E))|\ll |\Re(G^\nu_{0/1}(E))|$ is satisfied.  For this, we first apply the triangle inequality to each expression and obtain
\begin{align}\label{Eq_app:G0_TI}
    |\Im(G^\nu_0(E))|  &\leq 
    \frac{ \Gamma_\nu \delta_\nu}{2} \left|\frac{[1-\f (-i\delta_\nu+\mu_\nu)]}{z_0^2} \right| + \frac{1}{\beta_\nu} \left|    \sum_{k\ge0} \frac{1}{z_0^k} \G \Big(\frac{-i\pi(2k+1)}{\beta_\nu}+\mu_\nu \Big)  \right| + \frac{1}{2} \left|\G (\mp E) [1-\f (\mp E)] \right|
    \\\label{Eq_app:G1_TI}
    |\Im(G^\nu_1(E))|  &\leq 
    \frac{\Gamma_\nu \delta_\nu}{2} \left|\frac{\f (i\delta_\nu+\mu_\nu)}{z^1_1} \right| + \frac{1}{\beta_\nu}  \left| \sum_{k\ge0}  \frac{1}{z_1^k} \G \Big(\frac{i\pi(2k+1)}{\beta_\nu}+\mu_\nu \Big) \right| + \frac{1}{2} \left|\G (\mp E) \f (\mp E)  \right|
\end{align}
where we can identify the last term of each line as the real component of $G_{0/1}^\nu$ (see Eq.~\eqref{Eq_app:def_G0}-\eqref{Eq_app:def_G1}), i.e.
\begin{align}
    |\Im(G^\nu_0(E))|  &\leq 
    \frac{ \Gamma_\nu \delta_\nu}{2} \left|\frac{[1-\f (-i\delta_\nu+\mu_\nu)]}{z_0^2} \right| + \frac{1}{\beta_\nu} \left|    \sum_{k\ge0} \frac{1}{z_0^k} \G \Big(\frac{-i\pi(2k+1)}{\beta_\nu}+\mu_\nu \Big)  \right| + |\Re(G^\nu_0(E))| \\
    |\Im(G^\nu_1(E))|  &\leq 
    \frac{\Gamma_\nu \delta_\nu}{2} \left|\frac{\f (i\delta_\nu+\mu_\nu)}{z^1_1} \right| + \frac{1}{\beta_\nu}  \left| \sum_{k\ge0}  \frac{1}{z_1^k} \G \Big(\frac{i\pi(2k+1)}{\beta_\nu}+\mu_\nu \Big) \right| + |\Re(G^\nu_1(E))|
\end{align}
In the right-hand side fo the inequality we observe the scaling factor of the second and their term. For this we make use that $z^2_0$, $z^1_1$, $z_0^k$ and $z_0^1$ are linearly depended to $E$ which represents a generic $\e{k}{l}$ (see Eq.~\eqref{Eq_app:Sing_g0}-\eqref{Eq_app:Sing_g1}). Using the $\e{k}{l}= \tilde{\mu}-\omega(k-l)$ it is straightforward to see that $\e{k}{l} \sim \mathcal{N}$. We remind the reader that $\mathcal{N}$ is the size of the Hilbert space of the mechanics which tends to infinity theoretically. Therefore we can conclude that both the second and the third term of the first inequality scale as $1/\mathcal{N}$. Thus we can very loosely neglect the second and the third term resulting in 
\begin{align}
    |\Im(G^\nu_0(E))|  \leq 
    |\Re(G^\nu_0(E))|
    \qquad \text{and} \qquad
    |\Im(G^\nu_1(E))|  \leq |\Re(G^\nu_1(E))|
\end{align}
which is what we wanted to prove in this section. 

\subsection{Validity of secular approximation}\label{Appendix:Secular_aprox_cond}

In this section we study in detail under what conditions the secular proximation carried-out in Appendix~\ref{Appendix:LindblatForm} is valid, i.e., under what conditions is Eq.~\eqref{Eq_app:secular_aprox} satisfied. For clarity we re-write Eq.~\eqref{Eq_app:secular_aprox} below
\begin{equation}
    |\gamma^\nu_{k}(\omega_\alpha)| \ll |\omega_\alpha-\omega_{\alpha'}|  \quad \forall \, \omega_\alpha\neq \omega_{\alpha'}.
\end{equation}
On the one hand, if we explicitly express the left hand side of Eq.~\eqref{Eq_app:secular_aprox} for 
\begin{align}
    k=0: \quad\left|\gamma^\nu_{0}(\omega_\alpha) \right|
    &= \frac{1}{2} \underbrace{\left|\G(\omega_\alpha) \right|}_{\leq \Gamma_\nu} \underbrace{\left| [1-\f(\omega_\alpha)] \right|}_{\leq 1} \leq \frac{1}{2}\Gamma_\nu  \\ 
    k=1: \quad \left|\gamma^\nu_{1}(\omega_\alpha) \right| 
    &= \frac{1}{2} \underbrace{\left| \G(-\omega_\alpha)\right|}_{\leq \Gamma_\nu}  \underbrace{
    \left|\f(-\omega_\alpha) \right|}_{\leq 1}  \leq \frac{1}{2}\Gamma_\nu,
\end{align}
they can be bounded by the same value. Where we have used the $\f$ is the Fermi distribution and $\G$ is taken to be a singular spectral density in this work (see Eq.~\eqref{eq:spectral_Lorentz}). In the case of an alternative spectral density form, this bound must be modified. Thus Eq.~\eqref{Eq_app:secular_aprox} now results in 
\begin{equation}\label{Eq_app:secular_aprox_gamma}
    |\gamma^\nu_{k}(\omega_\alpha)| \leq \frac{1}{2}\Gamma_\nu \ll |\omega_\alpha-\omega_{\alpha'}|  \quad \forall \, \omega_\alpha\neq \omega_{\alpha'}.
\end{equation}
On the other hand, now we focus on the right hand side of Eq.~\eqref{Eq_app:secular_aprox_gamma}
\begin{align}
    |\omega_\alpha-\omega_{\alpha'}| 
    = \left| \omega_{\{ {\tau,j},{\sigma,l}\}}-\omega_{\{ {\tau',j'},{\sigma',l'}\}} \right|
    = \left| \tilde{E}_{l,\sigma}-\tilde{E}_{j,\tau} -(\tilde{E}_{l',\sigma'}-\tilde{E}_{j',\tau'}) \right|
    =\left|\omega \left(l-j-(l'-j')\right) \tilde{\mu} \left(\sigma-\tau-(\sigma'-\tau')\right) \right|. 
\end{align}
Under the condition $\omega_\alpha \neq \omega_{\alpha'}$, we can see that the smallest value possible is $\max\{\tilde{\mu},\omega\}$. Including this into condition detailed in Eq.~\eqref{Eq_app:secular_aprox_gamma} yields
\begin{equation}
    |\gamma^\nu_{k}(\omega_\alpha)| \leq \frac{1}{2} \Gamma_\nu \ll \max\{\tilde{\mu},\omega\} \leq |\omega_\alpha-\omega_{\alpha'}|  \quad \forall \, \omega_\alpha\neq \omega_{\alpha'}.
\end{equation}
Where it is straightforward to note that if we ask for $\Gamma_\nu \ll \max\{\tilde{\mu},\omega\} \, \forall \nu $ Eq.~\eqref{Eq_app:secular_aprox} is always satisfied. 

\section{Recovery of \qd~expressions}\label{Appendix:QD_recovery}

In this Appendix we showcase that we recover the typical \qd~expressions from Eq.~\eqref{Eq:ME_Redfield_final} of the main text when the \qd~+ \resonator is null, i.e. $\gw=0$. For this we first apply $\gw=0$ to the \textit{Redfield tensors} defined in Eq.~\eqref{Eq_app:A_nn}-\eqref{Eq_app:A_n-1n} of the main text resulting in
\begin{align}\nonumber
    \Rnn{n,}{n}{j}{m}{k}{l} & =  \frac{1}{2}  \Bigg[ 
    \delta_{n,0}
    \Big( R^{0\rightarrow 1}_\nu(\mu) \delta_{j,k} \delta_{ml} +  R^{0\rightarrow 1}_\nu(\mu) \delta_{l,m} \delta_{kj} \Big)  + \delta_{n,1}\Big( R^{1\rightarrow 0}_\nu(\mu) \delta_{j,k} \delta_{ml} + R^{1\rightarrow 0}_\nu(\mu) \delta_{l,m} \delta_{jk} \Big) \Bigg]\\ \label{Eq_app:A_nn}
    &= \delta_{n,0} \delta_{j,k} \delta_{ml} R^{0\rightarrow 1}_\nu(\mu)   + \delta_{n,1} \delta_{j,k} \delta_{ml} R^{1\rightarrow 0}_\nu(\mu)   \\ \label{Eq_app:A_n+1n}
    \Rnn{n+1,}{n}{j}{m}{k}{l} &= \frac{1}{2} \delta_{n,0} \delta_{j,k}\delta_{l,m} \Big( R^{1\rightarrow 0}_\nu(\mu)+ R^{1\rightarrow 0}_\nu(\mu)  \Big) =  \delta_{n,0} \delta_{j,k}\delta_{l,m} R^{1\rightarrow 0}_\nu(\mu) 
    \\ \label{Eq_app:A_n-1n}
    \Rnn{n-1,}{n}{j}{m}{k}{l} &= \frac{1}{2} \delta_{n,1} \delta_{j,k}\delta_{l,m}  \Big( R^{0\rightarrow 1}_\nu(\mu) + R^{0\rightarrow 1}_\nu(\mu) \Big) = \delta_{n,1} \delta_{j,k}\delta_{l,m} R^{0\rightarrow 1}_\nu(\mu) 
\end{align}
Note that the elements of the displacement operator yield delta functions as $\D_{ij}=\D^\dagger_{ij}=\delta_{ij}$ and $\e{k}{k}=\mu$ for this case that $\gw=0$. Replacing this into the projected Redfield equation Eq.~\eqref{Eq:ME_Redfield_final} of the main text yields
\begin{align}\nonumber
    \Pbra{j}\partial_t\trho_s^n \Pket{m} = -i\omega(j-m)\Pbra{j} \trho_s^n \Pket{m} 
    - \Big(\delta_{n,0}  R^{0\rightarrow 1}(\mu)  + \delta_{n,1} R^{1\rightarrow 0}(\mu) \Big) \Pbra{j} \trho_s^n \Pket{m} \\ \label{Eq_app:QD_equation_only_with_mec}
    + \delta_{n,0} R^{1\rightarrow 0}(\mu)  \Pbra{j} \trho_s^{n+1} \Pket{m}
     + \delta_{n,1} R^{0\rightarrow 1}(\mu)  \Pbra{j} \trho_s^{n-1} \Pket{m}. 
\end{align}
where we have simplified notation by using $\sum_\nu R^{0\rightarrow 1}_\nu(\mu) = R^{0\rightarrow 1}(\mu)$ and $\sum_\nu R^{1\rightarrow 0}_\nu(\mu) = R^{1\rightarrow 0}(\mu)$. Next we note that using that with $\gw=0$, the polaron transformation defied in Eq.~\eqref{eq:Polaron_trans} of the main text leaves the density matrix invariant, i.e. $e^{S}=\mathbf{1}$, $\trho_s=\rho_s=\rho_\text{\qd} \otimes \rho_\text{\resonator}$. Furthermore, 
$\Pbra{j}\trho^n_s\Pket{m} 
= \Fbra{j}{n}\trho_s \Fket{m}{n} 
= \Fbra{j}{n}\rho_\text{\qd} \otimes \rho_\text{\resonator} \Fket{m}{n} 
= \langle n| \rho_\text{\qd} |n \rangle \langle j | \rho_\text{\resonator} |m \rangle$ Now that the \qd~and \resonator density states are separable, to obtain the master equation corresponding only to the \qd, we trace out the \resonator~from Eq.~\eqref{Eq_app:QD_equation_only_with_mec} resulting in  
\begin{align}\label{Eq_app:QD_equation_only}
   \partial_t \rho^n_\text{\qd}  = 
    - \Big(\delta_{n,0}  R^{0\rightarrow 1}(\mu)  + \delta_{n,1} R^{1\rightarrow 0}(\mu) \Big) \rho^n_\text{\qd}  
    + \delta_{n,0} R^{1\rightarrow 0}(\mu)  \rho^{n+1}_\text{\qd} 
     + \delta_{n,1} R^{0\rightarrow 1}(\mu)  \rho^{n-1}_\text{\qd} . 
\end{align}
Where we have used
$ \sum_j \Pbra{j}\trho^n_s\Pket{j} 
= \langle n| \rho_\text{\qd} |n \rangle \sum_j \langle j | \rho_\text{\resonator} |j \rangle = \langle n| \rho_\text{\qd} |n \rangle = \rho^n_\text{\qd} 
$. Where Eq.~\eqref{Eq_app:QD_equation_only} are the the typical quantum dot rate equations with energy dependent tunneling rates~\cite{Timm2008,Potts2024}; showcasing that Eq.~\eqref{Eq:ME_Redfield_final} of the main text reduced to Eq.~\eqref{Eq_app:QD_equation_only} for $\gw=0$.

\section{Computation of classical analogues}\label{Appendix:Classical_analogues}

In this section we detail how we obtained the 'classical' analogues state $\rho_s^\text{diag}$ of the system used in Fig.~\ref{fig:Figure3}. The first step is to compute the diagonal elements of $\rho_s$ in the \NESS. This is done by solving the diagonal element of Eq.~\eqref{Eq:ME_Redfield_final} of the main text resulting in 
\begin{equation}
    \Pbra{m}\partial_t\trho_s^n \Pket{m} =  \sum_\nu \sum_{k}
    \Big( -\Rnn{n,}{n}{j}{m}{k}{k} \Pbra{k} \trho_s^n \Pket{k}
    + \Rnn{n+1,}{n}{j}{m}{k}{k} \Pbra{k} \trho_s^{n+1} \Pket{k}
     + \Rnn{n-1,}{n}{j}{m}{k}{k} \Pbra{k} \trho_s^{n-1} \Pket{k} \Big) =0 \quad \forall n,m.
\end{equation}
Note that this equation returns the \NESS~in the polaron basis $\trho_s^\text{NESS}$. The next step consists in transforming the obtained \NESS~into the original basis using Eq.~\eqref{eq:Polaron_trans} of the main text. This transformation may generate coherences in the \resonator~state. As we want to obtain a 'classical' state we only retain the diagonal elements of this state. Thus in the end the 'classical' analogous state of the system used in Sec.~\ref{Sec:Benchmark} is defined as
\begin{equation}
    \rho_s^\text{diag} = \text{diag}\Big( U^\dagger \trho_s^\text{NESS} U \Big).
\end{equation}

\section{Particle current derivation}\label{Appendix:I_computation}

In this section we carry out the computation of Eq.~\eqref{Eq:particle_current}. The first step is to transform de density matrix into the polaron picture as
\begin{align}
    \partial_t\langle \hat{N}_\nu \rangle = \partial_t \Tr\left(\hat{N}_\nu \rho\right) = \partial_t \Tr\left(e^{S}\hat{N}_\nu e^{-S}\trho\right) = \partial_t \Tr\left(\hat{N}_\nu \trho\right)
\end{align}
where in the last equality we have used that $[\hat{N}_\nu, S]= \sum_k [\cc\c, \gw(\bb-\b)\dd\d]=0$. Next we handle the time differential by switching to the Heisenberg representation denoted as $(\bullet)_\text{H}=e^{it\tilde{H}}(\bullet) e^{-it\tilde{H}}$:
\begin{align}\nonumber
    \partial_t\langle \hat{N}_\nu \rangle &= \partial_t \Tr\left(\hat{N}_\nu \trho\right) = \partial_t \Tr\left( (\hat{N}_\nu)_\text{H} (\trho)_\text{H}\right) =  \Tr\left( \partial_t(\hat{N}_\nu)_\text{H} (\trho)_\text{H}\right) = i\,\Tr\left( [\tilde{H}_H,(\hat{N}_\nu)_H] (\trho)_\text{H}\right) \\
    &=
    i\,\Tr\left( [\tilde{H}_s+H_\R+\tVr,\hat{N}_\nu] \trho\right) =
    i\, \Tr\left([\tVr,\hat{N}_\nu] \trho \right).
\end{align}
where in the end we returned to the Schr\"odinger picture. Now we separate the full density matrix under the polaron picture to the uncorrelated and the correlated using Eq.~\eqref{Eq_app:rho_un_correlated} and~\eqref{Eq_app:rho_correlated} 
\begin{align}\label{Eq_app:Partial_N_rho_circ}
    \partial_t\langle \hat{N}_\nu \rangle =
    i\, \Tr\left([\tVr,\hat{N}_\nu] \trho \right) = i\,\Tr\left( [\tVr,\hat{N}_\nu] \trho_\otimes \right) + i\, \Tr\left( [\tVr,\hat{N}_\nu] \trho_\circ \right) = i\, \Tr\left( [\tVr,\hat{N}_\nu] \trho_\circ \right) 
\end{align}
Where the first term vanished because the three-point functions that appear such as $\langle \cc \cc \c \rangle$ yield zero. Next, we use the solution for $\trho_{\circ}$ from Eq.~\eqref{Eq_app:tho_circ_sol_weak} in the Schr\"odinger picture, but before we convert it the the in the Schr\"odinger picture we first we apply the Markov approximation as it is applied in the interaction picture when the master equation was derived, see Eq.~\eqref{Eq_app:ME-local}. This yields Eq.~\eqref{Eq_app:tho_circ_sol_weak} as
\begin{equation}\label{Eq_app:rho_un_correlated_schro}
    \trho_{\circ}(t)= -i \int_0^{+\infty} \mathrm{d}s \, [e^{-i(\tHs+H_\R)s}\tVr e^{+i(\tHs+H_\R)s},\trho_{\otimes}(t)].
\end{equation}
where we took the change of variables $t'=t-s$. It is convenient at this point to define $\hat{O}(s)=e^{+i(\tHs+H_\R)s}$ for simplicity. Thus replacing Eq.~\eqref{Eq_app:rho_un_correlated_schro} into Eq.~\eqref{Eq_app:Partial_N_rho_circ} results in 
\begin{align}\label{Eq_app:Partial_N_integral}
    \partial_t\langle \hat{N}_\nu \rangle = \int_0^{+\infty} \mathrm{d}s \, \Tr\left( [\tVr,\hat{N}_\nu]  [\hat{O}(-s)\tVr \hat{O}(s),\trho_s(t) \otimes \rhoR] \right). 
\end{align}
From here, it is straightforward to note that $[\tVr,\hat{N}_\nu]=\sum_{k}  \t^* \S_1 \c - \t\cc\S_0$ using that the fermionic creation and annihilation operators of the reservoirs follow the anticommutation relations: $\{\hat{c}_{k',\nu'},\cc\}=\delta_{k,k'}\delta_{\nu,\nu'}$ and $\{\hat{c}^\dagger_{k',\nu'},\cc\}=\{\hat{c}_{k',\nu'},\c\}=0$. Therefore Eq.~\eqref{Eq_app:Partial_N_integral} results in 
\begin{equation}
    \partial_t\langle \hat{N}_\nu \rangle = \int_0^{+\infty} \mathrm{d}s \, \Tr\left( (\S_{1} \B_{\nu 1} - \B_{\nu 0}\S_{0})  [\hat{O}(-s)\tVr \hat{O}(s) ,\trho_s(t) \otimes \rhoR] \right) 
\end{equation}
where we have made use of the notation detailed in Eq.~\eqref{Eq_app:Bath_operators_int}. Now we explicitly obtain all four term as 
\begin{align}\nonumber
    \partial_t\langle \hat{N}_\nu \rangle
     &=  \int_0^{+\infty} \mathrm{d}s \, 
     \Tr\Big( 
     \S_1 \B_{\nu 1} \hat{O}(-s)\tVr \hat{O}(s) \trho_s(t) \rhoR \Big)
     - \Tr\Big( \S_1 \B_{\nu 1} \trho_s(t) \rhoR\hat{O}(-s)\tVr \hat{O}(s) \Big) \\ 
     & \qquad\qquad\quad
     - \Tr\Big( \B_{\nu 0} \S_0 \hat{O}(-s)\tVr \hat{O}(s) \trho_s(t) \rhoR\Big)
     +\Tr\Big( \B_{\nu 0} \S_0 \trho_s(t) \rhoR \hat{O}(-s)\tVr \hat{O}(s)
     \Big).
\end{align}
Next we express $\tVr$ by making use of Eq.~\eqref{eq:Coupling_expression} as $\tVr=\sum_{\nu'} (\S_0\B_{\nu' 1}^\dagger+\S_1\B_{\nu' 0}^\dagger)$and only retain terms that contribute to the correlation function of $C^\nu_{11}(s)$ and $C^\nu_{00}(s)$ , given that $C^\nu_{10}(s)$ and $C^\nu_{01}(s)$ are null (see Eq.~\eqref{Eq_app:C10}-\eqref{Eq_app:C01})
\begin{align}\nonumber
    \partial_t\langle \hat{N}_\nu \rangle
     &=  \sum_{\nu'} \int_0^{+\infty} \mathrm{d}s \, 
     \Tr\Big( 
     \S_1 \B_{\nu 1} \hat{O}(-s)\S_{0} \B_{\nu'1}^\dagger  \hat{O}(s) \trho_s(t) \rhoR \Big)
     - \Tr\Big( \S_1 \B_{\nu 1} \trho_s(t) \rhoR\hat{O}(-s)\S_{0} \B_{\nu'1}^\dagger  \hat{O}(s) \Big) \\ 
     & \qquad\qquad\quad
     - \Tr\Big( \B_{\nu 0} \S_0 \hat{O}(-s)\S_{1} \B_{\nu'0}^\dagger \hat{O}(s) \trho_s(t) \rhoR\Big)
     +\Tr\Big( \B_{\nu 0} \S_0 \trho_s(t) \rhoR \hat{O}(-s)\S_{1} \B_{\nu'0}^\dagger \hat{O}(s)
     \Big).
\end{align}
Now we separate the system subspace with the reservoir subspace using that the operators of each subspace commute, thus
\begin{align}\nonumber
    \partial_t\langle \hat{N}_\nu \rangle
     &=  \sum_{\nu'} \int_0^{+\infty} \mathrm{d}s \, 
     \Tr_s\Big( 
     \S_1 e^{-i\tHs s}\S_{0} e^{+i\tHs s} \trho_s(t)\Big) \Tr_\R\Big(\B_{\nu 1}  e^{-i H_\R s}  \B_{\nu'1}^\dagger  e^{+i H_\R s}  \rhoR \Big) \\ \nonumber
     & \qquad\qquad\quad
     - \Tr_s\Big(  \S_1 \trho_s(t) e^{-i\tHs s} \S_{0} e^{+i\tHs s}\Big)\Tr_\R\Big(   e^{-i H_\R s}  \B_{\nu'1}^\dagger  e^{+i H_\R s} \B_{\nu 1}  \rhoR \Big) \\ \nonumber
     & \qquad\qquad\quad
     - \Tr_s\Big( \S_0  e^{-i\tHs s} \S_{1} e^{+i\tHs s} \trho_s(t) \Big)\Tr_\R\Big( \B_{\nu 0}  e^{-i H_\R s}  \B_{\nu'0}^\dagger  e^{+i H_\R s}  \rhoR\Big) \\
     & \qquad\qquad\quad
     +\Tr_s\Big(  \S_0  \trho_s(t) e^{-i\tHs s} \S_{1} e^{+i\tHs s}\Big)\Tr_\R\Big( e^{-i H_\R s} \B_{\nu'0}^\dagger  e^{+i H_\R s}\B_{\nu 0}  \rhoR 
     \Big).
\end{align}
We have used the cyclical property of the trace among the reservoirs. Now the goal now is to express the traces along the reservoirs as correlation functions defined in Eq.~\eqref{Eq_app:C00} and~\eqref{Eq_app:C11}; 
\begin{align}\nonumber
    \partial_t\langle \hat{N}_\nu \rangle
     &= \int_0^{+\infty} \mathrm{d}s \, 
     \Tr_s\Big( 
     \S_1 e^{-i\tHs s}\S_{0} e^{+i\tHs s} \trho_s(t)\Big) C^\nu_{00}(s) 
     - \Tr_s\Big(  \S_1 \trho_s(t) e^{-i\tHs s} \S_{0} e^{+i\tHs s}\Big) C^\nu_{11}(-s) \\ 
     & \qquad\qquad\quad
     - \Tr_s\Big( \S_0  e^{-i\tHs s} \S_{1} e^{+i\tHs s} \trho_s(t) \Big) C^\nu_{11}(s) 
     +\Tr_s\Big(  \S_0  \trho_s(t) e^{-i\tHs s} \S_{1} e^{+i\tHs s}\Big)C^\nu_{00}(-s).
\end{align}
Now we explicitly express the traces over the system by using Eq.~\eqref{eq:Energy_system} 
\begin{align}\nonumber
    \partial_t\langle \hat{N}_\nu \rangle
     &= \sum_{j=0}^N\sum_{n=0,1} \int_0^{+\infty} \mathrm{d}s \, 
     \Fbra{j}{n}
     \S_1 e^{-i\tHs s}\S_{0} e^{+i\tHs s} \trho_s(t) \Fket{j}{n} C^\nu_{00}(s) 
     - \Fbra{j}{n} \S_1 \trho_s(t) e^{-i\tHs s} \S_{0} e^{+i\tHs s}\Fket{j}{n} C^\nu_{11}(-s) \\ 
     & \qquad\qquad\quad
     - \Fbra{j}{n}\S_0  e^{-i\tHs s} \S_{1} e^{+i\tHs s} \trho_s(t) \Fket{j}{n} C^\nu_{11}(s) 
     + \Fbra{j}{n} \S_0  \trho_s(t) e^{-i\tHs s} \S_{1} e^{+i\tHs s}\Fket{j}{n} C^\nu_{00}(-s),
\end{align}
and work out the reminder of the computation making use of Eq.~\eqref{Eq_app:S_0_elements}-\eqref{Eq_app:S_1_elements} for the projection of the system operator $\S_{0/1}$ and the integration results Eq.~\eqref{Eq_app:Int_C_start}-\eqref{Eq_app:Int_C_end} as 
\begin{align}\nonumber
    \partial_t\langle \hat{N}_\nu \rangle
     &= \sum_{j,i,l=0}^N
     \frac{1}{2} \G(\e{i}{l})[1-\f(\e{i}{l})] \D_{j,i} \DD_{i,l} \Pbra{l} \trho^{1}_s(t) \Pket{j}  
     + \frac{1}{2} \G(\e{j}{l})[1-\f(\e{j}{l})] \DD_{j,i} \D_{l,j} \Pbra{i}\trho^{1}_s(t) \Pket{l} \\
     & \qquad\qquad\quad
     - \frac{1}{2} \G(\e{l}{j})\f(\e{l}{j}) \D_{j,i} \DD_{l,j} \Pbra{i} \trho^{0}_s(t)\Pket{l}  
     - \frac{1}{2} \G(\e{l}{i})\f(\e{l}{i}) \DD_{j,i} \D_{i,l} \Pbra{l}\trho^{0}_s(t) \Pket{j}. 
\end{align}
Similar to Sec.~\ref{Appendix:ME}, we change the summation indexes in order to group the elements of the system density matrix
\begin{align}\nonumber
    \partial_t\langle \hat{N}_\nu \rangle
     &= \sum_{l,i,k=0}^N
     \frac{1}{2} \G(\e{i}{k})[1-\f(\e{i}{k})] \D_{l,i} \DD_{i,k} \Pbra{k} \trho^{1}_s(t) \Pket{l}  
     + \frac{1}{2} \G(\e{i}{l})[1-\f(\e{i}{l})] \DD_{i,k} \D_{l,i} \Pbra{k}\trho^{1}_s(t) \Pket{l} \\ 
     & \qquad\qquad\quad
     - \frac{1}{2} \G(\e{l}{i})\f(\e{l}{i}) \D_{i,k} \DD_{l,i} \Pbra{k} \trho^{0}_s(t)\Pket{l}  
     - \frac{1}{2} \G(\e{k}{i})\f(\e{k}{i}) \DD_{l,i} \D_{i,k} \Pbra{k}\trho^{0}_s(t) \Pket{l}. 
\end{align}
Making use of the definitions in Eq.~\eqref{def:QD_R01}-\eqref{def:QD_R10} of the main text we arrive to 
\begin{align}
    \partial_t\langle \hat{N}_\nu \rangle
     &= \sum_{l,i,k=0}^N 
     \frac{1}{2} \D_{l,i} \DD_{i,k}  \Big( R^{1\rightarrow 0}_\nu(\e{i}{k}) + R^{1\rightarrow 0}_\nu(\e{i}{l}) \Big) \Pbra{k}\trho^{1}_s(t) \Pket{l} 
     - \frac{1}{2} \D_{i,k} \DD_{l,i} \Big( R^{0\rightarrow 1}_\nu(\e{k}{i}) + R^{0\rightarrow 1}_\nu(\e{l}{i})  \Big)  \Pbra{k}\trho^{0}_s(t) \Pket{l}, 
\end{align}
where we identify the \textit{Redfield tensors} defined in Eq.~\eqref{eq:R_0_1} and Eq.~\eqref{eq:R_1_0} of the main text resulting in the tight expression
\begin{align}\label{Eq_app:partial_N_nu_final}
    \partial_t\langle \hat{N}_\nu \rangle
     &= \sum_{l,i,k=0}^N \Big( \Rnn{1}{0}{i}{i}{k}{l} \Pbra{k}\trho^{1}_s(t) \Pket{l} 
     -\Rnn{0}{1}{i}{i}{k}{l} \Pbra{k}\trho^{0}_s(t) \Pket{l}\Big). 
\end{align}
Replacing Eq.~\eqref{Eq_app:partial_N_nu_final} into the definition of the particle current given in Sec.~\ref{Sec:Current}
\begin{equation}
    \I =  e\, \partial_t\langle \hat{N}_\nu \rangle = e \sum_{l,i,k=0}^N \Big( \Rnn{1}{0}{i}{i}{k}{l} \Pbra{k}\trho^{1}_s(t) \Pket{l} 
     -\Rnn{0}{1}{i}{i}{k}{l} \Pbra{k}\trho^{0}_s(t) \Pket{l}\Big) ,
\end{equation}
we arrive to expression Eq.~\eqref{Eq:particle_current} of the main text. With this results we conclude the section. 

\section{Convergence of bath correlation functions}\label{Appendix:Bath_convergance}

\begin{figure}[ht]
    \centering
    \includegraphics[width=.8\linewidth]{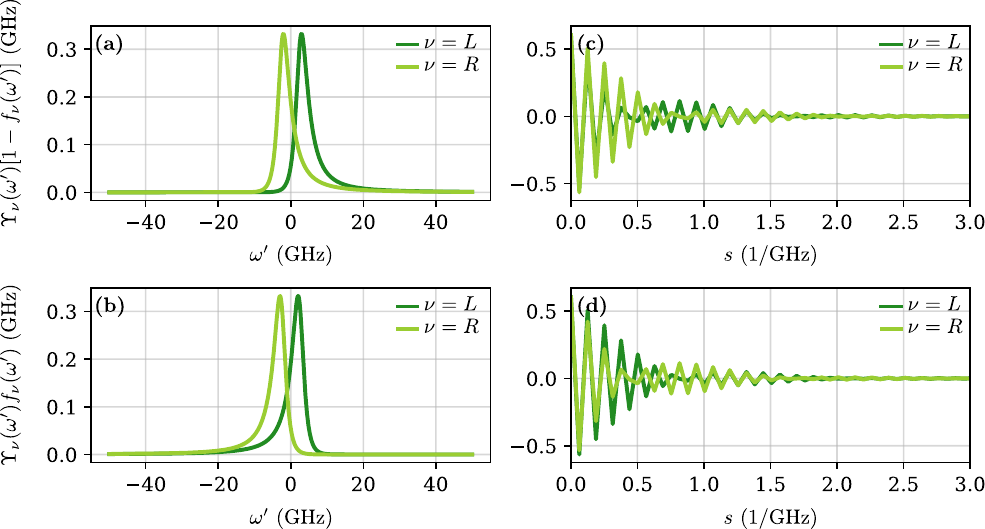}
    \caption{
    \textbf{(a)}-\textbf{(b)} Integration functions of the bath correlation functions as a function of frequency $\omega'$ 
    \textbf{(c)}-\textbf{(d)} Bath correlation functions detailed in Eq.~\eqref{Eq_app:C00} and \eqref{Eq_app:C11} respectively as a function of time $s$. 
    Parameters: $\Gamma_L/2\pi=\Gamma_R/2\pi=0.1$ GHz, $\omega/2\pi$= 1 GHz, $T_L=T_R=2$ GHz (corresponds to $15.28$ mK), $\gamma_L=-\gamma_R= 2.5$ GHz, $\delta_L=\delta_R=2$ GHz, and $\mu_L=-\mu_R=2.5$ GHz. }
    \label{fig:Figure2_appendix}
\end{figure}

In this appendix we come back to the discussion held in Sec.\ref{Sec:Note_dynamics} regarding the justification of the Markov approximation used in Eq.~\eqref{Eq:ME_Redfield_final} of the main text. This is always justified if the bath correlation time is smaller than any relevant time-scale of the system~\cite{Potts2024}. Though as we are only interested in the \textit{long time scales} of the system (the \NESS), to assume time-local it is sufficient to check that bath correlation functions, defined in Eq.~\eqref{Eq_app:C00}-\eqref{Eq_app:C11}, effectively decay at some time~\cite{Strasberg2022Book}. Here, we expose that for the parameters chosen in this work this condition is satisfied. In Fig.~\ref{fig:Figure2_appendix} we showcase an example case that represents the behavior of the parameters used in this work. In Figs.~\subfigref{fig:Figure2_appendix}{a} and~\subfigref{fig:Figure2_appendix}{b}  we plot the integration function of Eq.~\eqref{Eq_app:C00} and \eqref{Eq_app:C11} respectively. We integrate these functions numerically using the Qutip "spectrum\_correlation\_fft" function~\cite{Lambert2024Qutip5} to obtain the correlation function detailed in Eq.~\eqref{Eq_app:C00} and \eqref{Eq_app:C11}. These are plotted in Figs.~\subfigref{fig:Figure2_appendix}{c} and~\subfigref{fig:Figure2_appendix}{d} respectively. Note that after some oscillations, all the function decay to zero. Thus this justifies the use of the Redfield equation used to compute the \NESS~of the system for the parameters used in the paper.

\end{document}